\documentclass[twocolumn]{autart}

\usepackage{graphicx}      
\usepackage{listings}
\usepackage[utf8]{inputenc}
\usepackage{algorithm}
\usepackage{graphicx}
\usepackage{float}
\usepackage{fancyhdr}
\usepackage{enumitem}
\usepackage{tabularx}
\usepackage{amsmath}
\usepackage{verbatim}
\usepackage{amsfonts}
\usepackage{amssymb}
\usepackage{color}
\usepackage{xcolor,ifpdf}
\usepackage{enumitem}
\usepackage{multirow}
\usepackage{tipa}
\usepackage{enumitem}
\usepackage{hyphenat}
\graphicspath{{./Figures/}}

\newcommand{\beq}{\begin{equation}}
\newcommand{\eeq}{\end{equation}}
\newcommand{\beqa}{\begin{eqnarray}}
\newcommand{\eeqa}{\end{eqnarray}}
\newcommand{\beqan}{\begin{eqnarray*}}
\newcommand{\eeqan}{\end{eqnarray*}}

\newcommand{\sign}{\text{sign}}
\newcommand{\drivers}{\mathcal{K}}
\newcommand{\outputs}{\mathcal{O}}

\newcommand{\target}{\mathfrak{t}}
\newcommand{\targets}{\mathfrak{t}}
\newcommand{\source}{\mathfrak{s}}
\newcommand{\sources}{\mathfrak{s}}

\newcommand{\weights}{w}

\newcommand{\htwo}{{\mathcal{H}_2}}
\newcommand{\hinf}{{\mathcal{H}_\infty}}
\newcommand{\gainm}{1/(\left(I - A\right)^{-1})_{\source\target }}

\newcommand\Tstrut{\rule{0pt}{2.6ex}}         

\newcommand{\ErdosRenyi}{Erdős–Rényi }

\DeclareMathOperator{\trace}{Tr}

\usepackage{theorem}
\newtheorem{definition}{Definition}
\newtheorem{theorem}{Theorem}

\newtheorem{proposition}{Proposition}

\newtheorem{lemma}{Lemma}

{\theorembodyfont{\upshape}
	\newtheorem{remark}{Remark}

}

\newcommand{\tf}[2] {G^{(#1)}_{#2}}
\newcommand{\ir}[2] {g^{(#1)}_{#2}(t)}

\newcommand{\diag}[1] {\text{diag}#1}

\newcommand{\onevec}{\textbf{1}}
\usepackage{tikz}
\usetikzlibrary{shapes,arrows}
\usetikzlibrary{arrows,calc}

\def\qed{\hfill \vrule height 5pt width 5pt depth 0pt \medskip}
\newcommand{\proof}{\noindent {\bf Proof. }}

\begin{document}

\begin{frontmatter}

\title{Investigating the effect of edge modifications on networked control systems\thanksref{footnoteinfo}}

\thanks[footnoteinfo]{Work supported in part by a grant from the Swedish Research Council (grant n. 2015-04390). A preliminary version of this paper was presented at IFAC World Congress 2020 \cite{lindmarkEdgeMod2020}.}

\author[First]{Gustav Lindmark} 
\author[First]{Claudio Altafini} 

\address[First]{Division of Automatic Control, Dept. of Electrical Engineering, 
	Link\"oping University, SE-58183, Link\"oping, Sweden. email: {\tt\small gustav.lindmark@liu.se, claudio.altafini@liu.se}}

%

\date{}

%
%

\begin{abstract}
This paper investigates the impact of addition/removal of edges in a complex networked control system, for the purposes of improving its controllability, system performances or robustness to external disturbances.
The transfer function formulation we obtain allows to quantify the impact of an edge modification with the $\hinf$ and $\htwo$ norms. 
For stable networks with positive edge weights, we show that the $\hinf$ norm can be computed exactly for each possible single edge modification, as well as the associated stability margin. 
For the $\htwo$ norm we instead obtain a lower bound. 
Since this bound is linked to the trace of the controllability Gramian, it can be used for instance to reduce the energy needed for control. 
When instead the dynamics is of Laplacian type, then the norms become unbounded. However, the associated displacement  systems are stable and for them the effect of edge modifications can be quantified. 
In particular, in this case we provide an upper bound on the $ \hinf$ norm and compute the exact value of the $ \htwo $ norm for arbitrary edge additions.
\end{abstract}


\begin{keyword}   
Complex networks, Systems norms, Network topology design, Controllability, Robustness, Positive systems.
\end{keyword}

\end{frontmatter}


\section{Introduction}

The peculiar point of view of control theory on a networked system is that the network dynamics is influenced by external processes, which could be either driving inputs or known/unknown disturbances entering  the networks through  ``input nodes''.
Depending on the application, it might also be meaningful to specify ``output nodes'', corresponding to a subset of the nodes that are observed. 
The signals that constitute the driving inputs and the outputs depend on the specific application one is considering. 
For instance, in intelligent transportation systems inputs may correspond to variable speed limits or to traffic lights, and outputs to traffic flow measurements. In smart power grids, the actuators may correspond to the generators and the sensors to meters along the transmission lines. 
The presence of inputs and outputs gives rise to manyfold research problems which, while standard for the control community, carry a novel perspective when applied to network science. Among others, we can name controllability, observability, stability, control synthesis, performances, robustness, and so on. 
The classical approaches of network science are normally not enough to study these problems, but also the tools available to a control scientist have to be tailored and adapted to the large-scale network context. 
For instance, when dealing with controllability, one feature that is emerging is that the capacity of a node to steer the network depends on its surrounding topology and can  be characterized in terms of some centrality measure.
These centrality measures are often energy-related metrics based on the controllability Gramian (see, among others,  \cite{Chen2016Energy,altafiniLindmarkMinEnergy,pasqualetti2014controllability,yan2015spectrum}). 
Several methods have been proposed for placing a limited number of control actuators in the network in such a way to optimize (or improve) a given metric for control energy \cite{altafiniLindmarkMinEnergy,pasqualetti2014controllability,tzoumas2016minimal}.
When instead it comes to evaluating performances and robustness, the canonical system-theoretic approach is to make use of some system norm, like $\htwo$ or $\hinf$ norm, because they can capture how inputs impact the entire system by measuring the amplitude of its outputs. 
Several works have appeared in recent years trying to adapt these norms to the network context \cite{bamieh2012coherence,grunberg2018performance,hassan2017edge,patterson2010leader,pirani2018robustness,siami2013fundamental,siami2017growing,summers2015topology,young2010robustness}.
These works deal primarily with Laplacian dynamics and they try to characterize the impact of stochastic disturbances (equivalent, for linear systems, to impulsive inputs) on the so-called network coherence, i.e., the variance of the deviation from consensus. 
The coherence of a consensus network has been addressed for specific network topologies and dynamics, such as lattices \cite{bamieh2012coherence} trees \cite{siami2013fundamental,siami2017growing}, and oscillators \cite{grunberg2018performance}, as well as for general graphs \cite{young2010robustness}.

One of the original aspects of networked control systems is that often times extra degrees of freedom induced by the network structure of the system are available and can be leveraged for control design and/or performance evaluation. 
One such degree of freedom is linked to the choice of which nodes to control and observe.
The so-called driver node placement problem has been widely investigated for instance in the controllability and observability problems \cite{Chen2016Energy,altafiniLindmarkMinEnergy,pasqualetti2014controllability,yan2015spectrum}.
Another degree of freedom consists instead in rewiring the network by addition/\allowbreak removal/re-weighting of edges.
This approach is promising, given the significant impact that network topology has on control performances (see for instance \cite{bianchin2015role,parlangeli2011reachability}). 
Edge modifications are often feasible in applications and correspond to changes in e.g. connectivity of smart grids or traffic routing. 

When structural controllability (i.e., controllability based on presence/absence of edges only, not on edge weights \cite{lin1974structural}), is considered, graph-theoretic procedures can be used to identify a minimum number of edge additions that render a network controllable \cite{chen2018minimal,pichai1981vulnerability}.
Optimizing Gramian-based controllability metrics (i.e. minimizing the control energy) by edge modifications is however more difficult. 
Even when comparing edge modifications and control input placement as two different means to improve such metrics, edge modifications are generally more difficult.
This comes from the fact that the controllability Gramian as a function of the control inputs (columns in the input matrix $B$) is simpler than as a function of the edges (entries in the state update matrix $A$).
There are however a few studies in this direction: for a given budget of edges and weights that can be added, \cite{chanekar2019} applies differential analysis for maximization of the trace of the Gramian control energy metric.
In \cite{becker2017network}, re-weighting of existing edges is applied in order to reduce the worst case control energy as measured by the minimal eigenvalue of the Gramian.

Edge addition in consensus networks has received more attention, at least for continuous time network models  \cite{hagberg2008rewiring,hassan2017edge,siami2017growing,summers2015topology,zhang2017effect}. 
In this context, the focus is often on network robustness to external disturbances. Performance measures based on network coherence can be improved by the addition of new edges, and one interesting problem that is studied in e.g. \cite{siami2017growing,summers2015topology} is how to find the edges that give the largest gains.

In this paper we consider edge modifications in discrete-time linear networks with (\textit{i}) arbitrary stable dynamics, and (\textit{ii}) Laplacian dynamics.
For a given stable network with input and output nodes (in theory all nodes can be both input and output nodes) we derive a transfer function formulation for the changes in output 
caused by an edge modification.
The formulation is such that it can be applied in a large scale setting in which each of the $n^2 -n$ possible edges ($n$ is the total number of nodes) is considered for modification.

The focus in this paper is on networks with positive edge weights. For this important class (appearing often in applications) we use the theory of positive systems together with our transfer function formulation of the effects of edge modifications in order to derive several new results.
We show that the addition of an edge may render a stable network unstable if the weight is large and if new cycles appear in the network. The maximal weight by which an edge can be modified without causing instability can be computed explicitly. 
We also derive an analytical expression for the $\hinf$ norm of the transfer function for the differences in output due to the edge addition, and a lower bound for the $\htwo$ norm, both usable in a large scale network setting.
These norms can be interpreted as measures of the extent that an edge modification impacts the network and can have multiple applications. 
For instance while a large impact might represent a large risk when dealing with disturbance attenuation, it can instead represent an opportunity if we aim at improving controllability by reducing the control energy.
For example, the  $\htwo$ norm has a simple relationship with the trace of the controllability Gramian, hence adding the edges corresponding to the largest $\htwo$ norm is a way to reduce the energy needed for control.
The $ \hinf$ norm can instead be used to express the ``resilience'' of a network to edge modifications. In particular, the stability margins associated to these edge modifications can be computed exactly.
More importantly, the explicit expression we provide for the transfer function of an edge modification sheds light into the possible sources of instability of topology-based designs like the one we are proposing, namely, it shows that instability is triggered by the novel positive feedback loops created by the edge modification.

Since networks with Laplacian dynamics are marginally stable, their $\htwo$ and $\hinf$ norms are unbounded. However, their associated ``displacement systems'' are stable and provide meaningful interpretations of the effects of edge additions.
An exact calculation of the $ \hinf $ norm associated to an edge addition is no longer possible. 
However, we can provide an upper bound, the first we are aware of for the case of edge addition in general (undirected) graphs (some work exists on the $ \hinf$ performances evaluation on special structures, such as leader-follower \cite{Herman2015Nonzero,pirani2018robustness} or for edge addition on trees \cite{Pirani2019Graph}). 
We also show that the exact $\htwo$ norm of the displacement system of a network can be efficiently computed for all possible edge additions when there are inputs acting on all nodes. 
If these inputs are noises rather than control signals, then the $\htwo$ norm reflects the network coherence. Hence, we can for instance use our results to design edge additions that optimize such metric.

The rest of the paper is organized as follows.
In Section~\ref{sec:background}, definitions are given and the network model is presented.
In Section~\ref{sec:stableDynamics}, edge modifications in networks with arbitrary stable dynamics are considered, while Section~\ref{sec:LaplacianDynamics} deals with networks with Laplacian dynamics.
A few applications of the results are finally presented in Section \ref{sec:applications}.

A preliminary version of this paper was presented at IFAC World Congress 2020 \cite{lindmarkEdgeMod2020}. While this conference paper includes most of the results of Section \ref{sec:stableDynamics}, the material of Section \ref{sec:LaplacianDynamics} is presented here for the first time.

\section{Preliminaries}
\label{sec:background}

\subsection{Notation}

We denote by $\mathbb{R}^{n\times m}$ the set of $n\times m$ matrices with real valued entries. $\mathbb{R}^+$ is the set of non-negative real numbers, $\mathbb{N}$ the set of natural numbers and $\mathbb{N}_0$ the set of natural numbers including zero.
Given a matrix $M \in \mathbb{R}^{n\times m}$, let $M_{ij},\ i , j = 1,\dots,m,$ denote the element on row $i$ and column $j$.
We use $\bar{\sigma}(M)$ for the maximal singular value of $M$ and denote by $M^\dagger$ the pseudoinverse of $M$.
For $M$ and $N$ two matrices of the same dimension, $M\geq N$ should be interpreted element-vise, i.e. $M_{ij} \geq N_{ij}\ \forall i,j$. 
The spectral radius of the square matrix $M\in \mathbb{R}^{n\times n}$ is denoted by $\rho(M)$.	
The $i$-th vector of the canonical basis of $\mathbb{R}^n$ is denoted $e_i,\ i = 1,\dots, n$. Also, let $e_{ij} = e_i - e_j$, $i,j = 1,\dots,n,\ i \neq j$ and $  \onevec = \begin{bmatrix} 1 & \ldots & 1 \end{bmatrix}^\top$.

A graph is indicated by the triple $\mathcal{G} = (\mathcal{V},\mathcal{E},\mathcal{W})$, where $\mathcal{V} = \{1,\dots,n\}$ is the set of nodes, $\mathcal{E} \subseteq \{(i,j),\ i,j\in \mathcal{V} \}$ is the set of edges and $\mathcal{W} = \{w_{ji} \in \mathbb{R},\ i,j \text{ s.t. } (i,j) \in \mathcal{E}\}$ the set of edge weights.
The weighted adjacency matrix $\mathcal{A} \in \mathbb{R}^{n\times n} $ is defined in such a way that $\mathcal{A}_{ji} = w_{ji}$ if $(i,j) \in \mathcal{E}$ and $\mathcal{A}_{ji} =0$ otherwise.

A path in $\mathcal{G}$ is a subgraph of nodes  $\mathcal{V}^* = \{{i_1},\dots,{i_j}\}$ and  edges $\mathcal{E}^* = \{({i_1},{i_2}),\dots, ({i_{j-1}},{i_j})\}$. The path is directed from ${i_1}$ to ${i_j}$.
The cardinality of the set $\mathcal{S}$ is denoted by $|\mathcal{S}|$. 
For $\mathcal{S} = \{i_1,\dots, i_j\} \subseteq \{1,\dots,n\}$ we define $E_{\mathcal{S}} = [e_{i_1}\ \dots \ e_{i_j}] \in \mathbb{R}^{n\times |\mathcal{S}|}$.

For the vector $z\in \mathbb{R}^n$, $|z| = \sqrt{z^\top z}$ is its Euclidean norm. Given an input-output system $G$, we use $||G||_\htwo$ and $||G||_\hinf$ for its $\htwo$ resp. $\hinf$ norms.

\subsection{Network model}

Consider a network represented by the graph $\mathcal{G} = (\mathcal{V},\mathcal{E},\mathcal{W})$. 
Each external input is assumed to act only on one node which is then called an \emph{input node}. The set of input nodes is $\drivers \subseteq \mathcal{V}$, $|\mathcal{K}| = n_\mathcal{K}$. Similarly, the \emph{output nodes} are given by the set $\mathcal{O}\subseteq \mathcal{V}$, $|\mathcal{O}| = n_\mathcal{O}$. Observe that $\drivers = \mathcal{V}$ and/or $\outputs = \mathcal{V}$ is possible.
We consider the following discrete-time, linear, time-invariant model of the network dynamics
\begin{align}
	\begin{array}{lll}
	x(t+1) &= &Ax(t) + Bu(t),\\
	y(t) &= &Cx(t),
	\end{array}
	\label{eq:discreteStateUpdate}
\end{align}
where $x(t) \in \mathbb{R}^n$ is the state of the network at time $t\in \mathbb{N}_0$.
For the directed networks with arbitrary stable dynamics that are studied in Section \ref{sec:stableDynamics}, the state update matrix is simply the adjacency matrix, i.e. $A = \mathcal{A}$. The undirected networks with Laplacian dynamics considered in Section \ref{sec:LaplacianDynamics} has however a slightly different expression for $A$ (see \eqref{eq:undirGraphLapFromEdges}-\eqref{eq:ConsensusNw} below).
In \eqref{eq:discreteStateUpdate},
$B=E_\drivers \in \mathbb{R}^{n\times n_\drivers}$ is the input matrix, $u(t) \in \mathbb{R}^{n_\drivers}$ is the input vector, $C= E_\outputs^\top \in \mathbb{R}^{n_\outputs \times n}$ is the output matrix, and $y(t) \in \mathbb{R}^{n_\outputs}$ the output vector.

In this paper we sometimes consider input/output relations between other sets of nodes than $\drivers$ and $\outputs$.
To make the presentation clear, we introduce the following notation: for two sets of nodes, $\mathcal{S}_1 \subseteq \mathcal{V}$ and $\mathcal{S}_2 \subseteq \mathcal{V}$, the transfer function from inputs applied to $\mathcal{S}_1$ to the states of $\mathcal{S}_2$ (intended as output nodes) is denoted by
\begin{align}
	\tf{A}{\mathcal{S}_2 \mathcal{S}_1} = 
	\left[\begin{array}{c|c}
	A & E_{\mathcal{S}_1} \\ 
	\hline 
	E_{\mathcal{S}_2}^\top & 0 \Tstrut
	\end{array}\right].
	\label{eq:networkSystemStdForm1}
\end{align}
Moreover, the impulse response is
\begin{align}
\ir{A}{\mathcal{S}_2 \mathcal{S}_1} = \begin{cases}0 \text{ for } t=0,\\
E_{\mathcal{S}_2}^\top A^{t-1} E_{\mathcal{S}_1} \text{ for } t \in \mathbb{N}.\end{cases}
\label{eq:networkSystemStdForm2}
\end{align}
With this notation we can write the system \eqref{eq:discreteStateUpdate} as
\begin{align*}
	&y(t) = \tf{A}{\outputs \drivers} u(t),\ \tf{A}{\outputs \drivers} = \left[\begin{array}{c|c}
	A & B \\ \hline C & 0 \Tstrut
	\end{array}\right].
\end{align*}

All the networks considered in this paper have positive edge weights.
\begin{definition}
	The linear system $(A,B,C)$ is said to be \emph{externally positive} if its forced output is non-negative for every non-negative input function. It is said to be \emph{positive} if for every non-negative initial state and for every non-negative input, both its state and outputs are non-negative. \label{def:posSystem}
\end{definition}
Clearly, positivity implies external positivity. A necessary and sufficient condition for $(A,B,C)$ being externally positive is that the impulse response is non-negative. Moreover, $(A,B,C)$ is positive if and only if $A\geq 0$, $B\geq 0$ and $C\geq 0$ \cite{farina2011positive}. 

\subsection{System norms and network centrality measures}

Internal stability of a system on the form \eqref{eq:discreteStateUpdate} holds true if $\rho(A) < 1$, while the system is marginally stable if $\rho(A) = 1$ is a simple eigenvalue of $A$.
The networks considered in Section \ref{sec:stableDynamics} are assumed internally stable, while the Laplacian systems of Section~\ref{sec:LaplacianDynamics} are marginally stable.

For an arbitrary internally stable discrete-time LTI system \\ $G~=~\left[\begin{array}{c|c}
A & B \\ \hline C & 0 \Tstrut
\end{array}\right]$ with impulse response $g(t)~=~CA^{t-1}B\in\mathbb{R}^{n_\outputs \times~n_\drivers},\ t \in \mathbb{N}$, the $\htwo$ norm is given by
\begin{align}
	||G||_\htwo^2 &
	= \sum_{t=1}^\infty \sum_{\substack{ i=1,\dots,n_{\drivers}, \\ j =1,\dots, n_\outputs}} \left( g_{ji}(t) \right)^2
	=\sum_{t=1}^\infty \trace g^\top(t) g(t),
	\label{eq:H2normGeneral}
\end{align}
which can be equivalently written as
\begin{align*}
	& ||G||_\htwo^2 = \trace \left(C W C^\top \right), \text{ where} \\
	& W = \sum_{\tau= 0}^\infty A^\tau B B^\top (A^\top)^\tau 
\end{align*}	
is the (infinite time) controllability Gramian.

For network systems of the form \eqref{eq:networkSystemStdForm1}-\eqref{eq:networkSystemStdForm2}, the expression \eqref{eq:H2normGeneral} can be rewritten as follows.
First, with the single input node $i\in \mathcal{V}$ and the single output node $j\in \mathcal{V}$, 
\begin{align*}
	||\tf{A}{ji}||_\htwo^2 &= 
	\sum_{t=1}^\infty \left( e_j^\top A^{t-1} e_i\right)^2
	= \sum_{\tau=0}^\infty \left((A^\tau)_{ji}\right)^2 \nonumber := \varepsilon_{i \to j}.
\end{align*}
The quantity $\varepsilon_{i \to j}\geq 0$ was introduced in \cite{lindmark2019combining} and referred to as the \emph{walk energy} from node~$i$~to~node~$j$. 
In the multiple-input-multiple-output (MIMO) case, with $\mathcal{S}_1,\ \mathcal{S}_2 \subseteq \mathcal{V}$ arbitrary, straightforward calculations give 
\begin{align}
	||\tf{A}{\mathcal{S}_2 \mathcal{S}_1}||^2_\htwo 
	&= \sum_{\substack{ \forall i\in \mathcal{S}_1, \\ \forall j \in \mathcal{S}_2}} ||\tf{A}{ji}||^2_\htwo.
	\label{eq:H2normFromGramian_1a}
\end{align}

The following two network centrality metrics are variants of the ones proposed in \cite{lindmark2019combining}:
\begin{definition}
	The {\em input-to-node} resp. {\em node-to-output} network centralities are given by
	\begin{align*}
	&q_i = ||\tf{A}{i \drivers}||_\htwo^2 = \sum_{\forall j \in \mathcal{K}} \varepsilon_{j \to i},\ i \in \mathcal{V}, \\
	&p_i = ||\tf{A}{\outputs i}||_\htwo^2 = \sum_{\forall j \in \mathcal{O}} \varepsilon_{i \to j},\ i \in \mathcal{V}.
	\end{align*}
	\label{def:piqi}
\end{definition}
The walk energies and the input-to-node resp. node-to-output centrality measures can be computed for all nodes even in large scale networks by solving Lyapunov equations.
In this paper we shall see that these quantities are important for the impact of edge-modifications in the network.

The $\hinf$ norm of an LTI system is induced by the $\mathcal{L}_2$ signal norm,
\begin{align}
	||G||_\hinf = 
	\sup_{u(t)} \frac{||G u(t)||_{\mathcal{L}_2}}{||u(t)||_{\mathcal{L}_2}} = 
	\sup_{\theta} \ \bar{\sigma}(G(\theta)),
	\label{eq:HinfDef}
\end{align}
where $G(\theta),\ \theta \in [-\pi,\pi]$ is the frequency function.
While the norm cannot usually be computed directly for LTI systems (rather, one has to test if $||G||_\infty < \gamma$ for some $\gamma > 0$),
for positive systems the following proposition can be used:
\begin{proposition}[\cite{tanaka2011bounded}]
	Let $G(\theta)$ be the frequency function of a stable externally positive LTI system. Then 
	\begin{enumerate}
		\item $G(0) \geq 0$,
		\item $||G||_\hinf = \bar{\sigma}(G(0))$.
		\label{eq:posHinfGain}
	\end{enumerate}
	\label{prop:posProp2}
\end{proposition}
That is, the $\hinf$ norm of a stable externally positive system coincides with the spectral norm of the steady state transfer function.
For the network model \eqref{eq:networkSystemStdForm1}, with $A\geq 0,\ \rho(A) < 1$ and the sets $\mathcal{S}_1,\mathcal{S}_2 \subseteq \mathcal{V}$, the steady state transfer function is
\begin{align*}
	\tf{A}{\mathcal{S}_2 \mathcal{S}_1}(0) &= E_{\mathcal{S}_2}^\top \left( I + A + A^2 + ... \right) E_{\mathcal{S}_1}  \nonumber \\
	&= E_{\mathcal{S}_2}^\top ( I-A )^{-1} E_{\mathcal{S}_1},
\end{align*}
which makes the exact computation of the $\hinf$ norm easy.

The following result will be used to determine the internal stability of positive systems.
\begin{proposition}{\cite{farina2011positive}}
	For $A\geq 0$, $(I-A)^{-1}$ exist and is non-negative if and only if $\rho(A) < 1$.
	\label{prop:posProperties2}
\end{proposition}

\section{Edge modifications in networks with stable dynamics}
\label{sec:stableDynamics}

Consider a network given by the state update matrix $A = \mathcal{A}$ and the sets $\mathcal{K}$ and $\mathcal{O}$. 
Assume that the edge $(\source,\target)$, $\source,\target \in \mathcal{V}$, is modified with the weight $w$ such that the state update matrix of the modified network becomes
\begin{align}
	\bar{A}	 &= A + e_{\target} w e_\source^\top.
	\label{eq:edgeModification}
\end{align}
We always assume that $w \geq -A_{\target \source}$ so that $\bar{A} \geq 0$ (i.e. the modification preserves network positivity). 
In the following, we use the triplet  $\{(\sources,\targets),\weights\}$ to identify the modification. 

Denote $y(t) = \tf{A}{\outputs \drivers}u(t)$ and $\bar{y}(t) = \tf{\bar{A}}{\outputs \drivers} u(t)$ the outputs of the networks associated to $A$ and $\bar{A}$. 
For a given input sequence $u(t),\ t=0,1,\dots$, the difference 
\begin{align}
	y^\delta(t) &= \bar{y}(t) - y(t) = \left(\tf{\bar{A}}{\outputs \drivers} - \tf{A}{\outputs \drivers}\right)u(t) \label{eq:GdeltaDef}
\end{align}
is the change in the states of the output nodes due to the edge modification \eqref{eq:edgeModification}.
The corresponding transfer function,
\begin{align}
	G^\delta = \tf{\bar{A}}{\outputs \drivers} - \tf{A}{\outputs \drivers},
	\label{eq:GdeltaAltDef}
\end{align}
is from now on referred to as the \emph{delta system}.

\begin{proposition}
	Consider a network with adjacency matrix $A\geq 0$, sets $\drivers,\ \outputs$, and a modification $\{(\sources,\targets),\weights\}$.
	If the weight $w>0$, then $G^\delta$ is externally positive. If instead $-A_{\target \source} \leq w < 0$, then $-G^\delta$ is externally positive.
\end{proposition}

\proof
	With a non-negative input sequence $u(t)$,
	\begin{align*}
	&w > 0 \Rightarrow \bar{A} \geq A \geq 0 \Rightarrow \bar{A}^{t-1} \geq A^{t-1} \geq 0\  \forall t \in \mathbb{N} \\
	& \Rightarrow y^\delta(t) = G^\delta u(t) = C(\bar{A}^{t-1} - A^{t-1})B u(t) \geq 0\  \forall t \in \mathbb{N},
	\end{align*}
	i.e. the output of $G^\delta$ is non-negative. Hence, $G^\delta$ is externally positive by definition.
	
	On the other hand, choosing $w$ s.t. $-A_{\target \source} < w < 0$ means reducing the weight of the existing edge $(\source,\target)$, and choosing $w = -A_{\target \source}$ means completely removing it. With $u(t)$ non-negative, we obtain
	\begin{align*}
	A \geq \bar{A} \geq 0 \Rightarrow -y^\delta(t) = -G^\delta u(t) \geq 0 \  \forall t \in \mathbb{N},
	\end{align*}
	i.e. $-G^\delta$ is externally positive.
	 	
\qed

We seek an expression for the delta system that depends explicitly on $\sources,\targets$ and $\weights$, but not on $\bar{A}$.
\begin{proposition}
	It holds
	\begin{align}
		G^\delta = \tf{A}{\outputs \targets}  \left( I - \weights\tf{A}{\sources \targets} \right)^{-1} \weights \tf{A}{\sources\drivers}.
		\label{eq:Gdelta}
	\end{align}
	\label{prop:deltaSystem}
\end{proposition}

\proof
	We have
	\begin{align*}
		G^\delta =  
		\left[\begin{array}{c|c}
		\bar{A} & B \\ \hline C & 0 \Tstrut
		\end{array}\right] - 
		\left[\begin{array}{c|c}
		A & B \\ \hline C & 0 \Tstrut
		\end{array}\right]
		= \left[\begin{array}{cc|c}
		\bar{A} & 0 & B \\ 0 & A  & B \\ \hline C & -C & 0 \Tstrut
		\end{array}\right].
	\end{align*}
	The formulation above corresponds to the state vector $[\bar{x}^\top x^\top]^\top$, where $\bar{x}$ is the state of the modified network and $x$ that of the original network.
	Define the state transformation 
	\begin{align*}
		&\begin{bmatrix}	\tilde{x} \\ x \end{bmatrix} = \begin{bmatrix} \bar{x} - x \\ x \end{bmatrix} = \begin{bmatrix} I & -I \\ 0 & I \end{bmatrix} \begin{bmatrix} \bar{x} \\ x \end{bmatrix},\ \text{with inverse} \\
		&\begin{bmatrix}	\bar{x} \\ x \end{bmatrix} = \begin{bmatrix} I & I \\ 0 & I \end{bmatrix} \begin{bmatrix} \tilde{x} \\ x \end{bmatrix}.
	\end{align*}
	Changing basis, 
	\begin{align*}
		&\left[\begin{array}{cc|c}
		I & -I & 0 \\ 0 & I & 0 \\ \hline 0 & 0 & I \Tstrut
		\end{array}\right]
		\left[\begin{array}{cc|c}
		\bar{A} & 0 & B \\ 0 & A  & B \\ \hline C & -C & 0 \Tstrut
		\end{array}\right]
		\left[\begin{array}{cc|c}
		I & I & 0 \\ 0 & I & 0 \\ \hline 0 & 0 & I \Tstrut
		\end{array}\right] \nonumber \\
		&=
		\left[\begin{array}{cc|c}
		\bar{A} & e_\targets \weights e_\sources^\top  & 0 \\ 0 & A & B \\ \hline C & 0 & 0 \Tstrut
		\end{array}\right] 
		=
		\left[\begin{array}{c|c}
		\bar{A} & e_\targets \\ \hline C &  0 \Tstrut
		\end{array}\right]
		\weights
		\left[\begin{array}{c|c}
		A & B \\ \hline e_\sources^\top &  0 \Tstrut
		\end{array}\right], \nonumber
	\end{align*}
	from which
	\begin{align}
		G^\delta &= \tf{\bar{A}}{\outputs \targets} \weights \tf{A}{\sources \drivers}.
		\label{eq:tfDifference1}
	\end{align}
	Analogous calculations give
	\begin{align*}
		&\tf{\bar{A}}{\outputs \targets} - \tf{A}{\outputs \targets} = \tf{\bar{A}}{\outputs \targets} \weights \tf{A}{\sources \targets} \\
		& \Leftrightarrow \tf{\bar{A}}{\outputs \targets} = \tf{A}{\outputs \targets} \left( 1 - \weights \tf{A}{\sources \targets} \right)^{-1},
	\end{align*}
	which together with \eqref{eq:tfDifference1} gives \eqref{eq:Gdelta}.  
\qed

Considering \eqref{eq:GdeltaDef}, it is the input-output relation of $G^\delta$ that is of interest rather than the states.
However, we can observe that $G^\delta$ can always be realized with $2n$ states. That is, two states for each node $i \in \mathcal{V}$ are sufficient: one state which is the same as in the original network (i.e. $x_i$) and another which corresponds to the  difference w.r.t. the modified network (i.e. $\bar{x}_i - x_i$), see the proof of Proposition \ref{prop:deltaSystem}.
\tikzset{
	block/.style = {draw, rectangle, 
		minimum height=.7cm, 
		minimum width=1.2cm},
	input/.style = {coordinate,node distance=.8cm},
	output/.style = {coordinate,node distance=.8cm},
	myOut/.style = {coordinate,node distance=1.6cm},
	arrow/.style={draw, -latex,node distance=1.6cm},
	pinstyle/.style = {pin edge={latex-, black,node distance=1.6cm}},
	sum/.style = {draw, circle, node distance=1.6cm}
}
\begin{figure}[ht]
	\begin{center}
		\begin{tikzpicture}[auto, node distance=1.6cm,>=latex',remember picture]
		\node [input, name=input] {};
		\node [block, right of=input] (G1) {$\tf{A}{\sources\drivers}$};
		\node [sum, right of=G1] (sum) {\footnotesize $+$};
		\node [block, right of=sum] (delta) {$\weights$};
		\node [output, right of=delta] (output) {};
		\node [block, right of=output] (G2) {$\tf{A}{\outputs \targets}$};
		\node [myOut, right of=G2] (myOut) {};
		\node [block, below of=delta] (feedback) {$\tf{A}{\sources\targets}$};
		\draw [draw,->] (input) -- node {$u(t)$} (G1);
		\draw [->] (G1) -- node {} (sum);
		\draw [->] (sum) -- node {} (delta);
		\draw [-] (delta) -- node {} (output);
		\draw [->] (output) -- node [name=y] {}(G2);
		\draw [->] (G2) -- node {}(myOut);
		\draw [->] (y) |- node [above,pos=0.79] {} (feedback) ;
		\draw [->] (feedback) -| node[pos=1.12] {} node [near end] {} (sum);
		\draw (2.8, -2.1) rectangle (6.3, 1) [dashed];
		\node[] at (6,.7) {$G_c^\delta$};
		\end{tikzpicture}
	\end{center}
	\caption{Block diagram of the delta system $G^\delta$.}\label{fig}
	\label{fig:deltaSystem}
\end{figure}
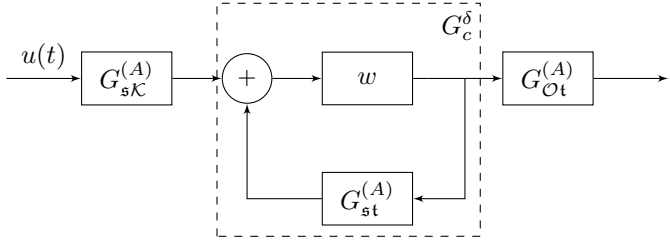

Figure \ref{fig:deltaSystem} is a block diagram illustration of the delta system \eqref{eq:Gdelta}.
We can view $G^\delta$ in \eqref{eq:Gdelta} as composed of three parts through which the edge modification \eqref{eq:edgeModification} perturbs the network transfer function.	
The first and last parts are the transfer functions $\tf{A}{\sources \drivers}$ and $\tf{A}{\outputs \targets}$ respectively. 
If $\nexists$ at least one path from a node $k \in \drivers$  to $\sources$, then $\tf{A}{\sources\drivers}$ is identically zero since $\ir{A}{\sources\drivers} = e_\sources^\top A^t E_\drivers = 0\ \forall t$. Similarly, if $\nexists$ at least one path from $\targets$ to any $o \in \outputs$, then $\tf{A}{\outputs \targets}$ is zero. 
In both these cases the delta transfer function is zero.
Intuitively, we can extend this reasoning to the transfer functions $\tf{A}{\sources\drivers}$, $\tf{A}{\outputs \targets}$ being ``small'': if $\tf{A}{\sources\drivers}$ is small then the source node of the new edge is essentially unaffected by the inputs, while if $\tf{A}{\outputs \targets}$ is small the target node does not affect the outputs.
The middle part of $G^\delta$ is
\begin{align*}
	G^\delta_c = \left(1 - \weights \tf{A}{\sources \targets} \right)^{-1}\weights,
\end{align*}
i.e. the new edge and $\tf{A}{\sources \targets}$ in closed positive feedback loop. 
Edge modification/addition for control of large scale networks is studied with differential analysis in \cite{chanekar2019}. 
Such analysis is only valid for the weight $\weights$ small, corresponding to the approximation $G^\delta \approx \tf{A}{\outputs \targets} \weights \tf{A}{\sources \drivers}$.
However, as $\weights$ increases, the effect of the feedback loop quickly becomes significant, and it cannot be neglected.


With the delta system it is possible to characterize exactly the implications of a specific edge modification $\{(\sources,\targets),\weights\}$.
Here we present results for the case when each of the $n(n-1)$ possible edges $(\source,\target)$ in a positive large scale network have to be considered for modification.
The results are made possible by the fact that the formulation \eqref{eq:Gdelta} does not depend explicitly on $\bar{A}$. This allows computationally heavy operations to be performed only once, and to reuse the results for the analysis of each single edge.

\subsection{Stability bounds and the $\hinf$ norm of $G^\delta$}

Due to the feedback loop in the delta system, an edge modification may render a stable network system  unstable.
For continuous-time positive linear systems, the related issue of internal stability to structured perturbations has been studied e.g. in \cite{son1996robust}. The following theorem investigates the problem for networked systems and edge modifications.
\begin{theorem}
	Consider a network with adjacency matrix $A\geq 0$, $\rho(A)<1$, and the edge modification $\{(\source,\target),w\},\ w > 0$.
	If the original network has no path $\target \to \source$, then the modified network is internally stable for any $w >0$. 
	On the other hand, if there is a path $\target \to \source$, then the modified network is internally stable if and only if $0 \leq w < 1/\left((I-A)^{-1}\right)_{\source \target}$.
	\label{prop:stabRad}
\end{theorem}
\proof
	First note that for $A\geq 0,\ \rho(A)< 1$, 
	\begin{align*}
		e_\source^\top (I-A)^{-1} e_\target  &= e_\source^\top \left(I + A + A^2 + \dots \right)  e_\target \\ 
	 &\begin{cases}
				= 0 \text{ if } \nexists \text{ a path } \target \to \source, \\
				> 0 \text{ if } \exists \text{ a path } \target \to \source.
			\end{cases}
	\end{align*}
	By Proposition \ref{prop:posProperties2}, $\rho(\bar{A}) < 1$ holds true if and only if \break $(I-\bar{A})^{-1} \geq 0$. We will show that this is the case for the weights specified by the theorem.		
	
	\emph{Sufficiency:} Using the matrix inversion lemma \cite{horn2012matrix}, we obtain
	\begin{align}
		(I-\bar{A})^{-1} &= (I - A - e_\target w e_\source^\top)^{-1} \nonumber \\
		&= (I - A)^{-1} + 
	 \frac{(I - A)^{-1} e_\target w e_\source^\top (I - A)^{-1}}{1 - w e_\source^\top (I-A)^{-1} e_\target}. \label{eq:midResult}
	\end{align}
	Given that $(I-A)^{-1}\geq 0$, the expression \eqref{eq:midResult} is non-negative if the denominator is s.t.
	\begin{align*}
		1 -w e_\source^\top (I-A)^{-1}e_\target > 0.
	\end{align*}
	This is true for any $w > 0$ if $\nexists$ a path $\target \to \source$, and for $0 < w < 1/\left((I-A)^{-1}\right)_{\source \target}$ if $\exists$ a path $\target \to \source$.
	
	\emph{Necessity:} 
	Assume that $\exists$ a path $\target \to \source$ and, by contradiction, that $ w > 1/\left((I-A)^{-1}\right)_{\source \target} $. Then 
	\begin{align*}
		 &0 > 1-w\left((I-A)^{-1}\right)_{\source \target} > -w\left((I-A)^{-1}\right)_{\source \target}\\
		\Rightarrow\  &\frac{1}{1 - w\left((I-A)^{-1}\right)_{\source \target}} < \frac{1}{-w\left((I-A)^{-1}\right)_{\source \target}}.
	\end{align*}
	The element on row $s$ and column $\target$ of \eqref{eq:midResult} is
	\begin{align*}
		((I-&\bar{A})^{-1})_{\source \target} \\
		&=\left((I-A)^{-1}\right)_{\source \target} + \frac{\left((I-A)^{-1}\right)_{\source \target} w \left((I-A)^{-1}\right)_{\source \target}}{1 -w\left((I-A)^{-1}\right)_{\source \target}} \\
		&<\left((I-A)^{-1}\right)_{\source \target} + \frac{\left((I-A)^{-1}\right)_{\source \target} w \left((I-A)^{-1}\right)_{\source \target}}{-w\left((I-A)^{-1}\right)_{\source \target}} \\
		&= 0,
	\end{align*}
	i.e. it does \emph{not} hold $(I-\bar{A})^{-1} \geq 0$, hence from Proposition \ref{prop:posProperties2} the modified network is \emph{not} stable. Finally, with $w = 1/\left((I-A)^{-1}\right)_{\source \target}$, the inverse $(I-\bar{A})$ does not exist. This case corresponds to $\bar{A}$ marginally stable.
\qed

When the conditions on the weight $w$ of Theorem \ref{prop:stabRad} are met, then both $\tf{A}{\outputs \drivers}$ and $\tf{\bar{A}}{\outputs \drivers}$ are stable. Hence, also $G^\delta = \tf{\bar{A}}{\outputs \drivers} - \tf{A}{\outputs \drivers}$ is stable and $||G^\delta||_\htwo$, $||G^\delta||_\hinf$ are bounded.

We can interpret Theorem \ref{prop:stabRad} in terms of cycles in the network: $\tf{A}{\source \target} > 0$ if there is a path $\target \to \source$. This path forms a cycle with the modified edge $\{(\source,\target),w\}$.
As a consequence, only edge additions that create new cycles, or edge modifications that increase the weight of an existing edge that is part of a cycle, may cause instability.
Edge removal or reduction of the weight of an existing edge in a positive network will on the other hand never cause instability \cite[p. 43]{farina2011positive}.
To see this, consider \eqref{eq:edgeModification} with $-A_{\target \source} \leq w < 0$ implying that $A \geq \bar{A} \geq 0$. Let $\bar{x}_e(t)$ resp. $x_e(t)$ denote the free motion of the modified resp. original network, then with $\bar{x}_e(0) = x_e(0)$ it follows that $x_e(t) \geq \bar{x}_e(t) \geq 0$ $\forall t$.\\
	
The $\hinf$ norm of $G^\delta$ can be computed exactly.
\begin{theorem}
	Consider a network with adjacency matrix $A\geq 0,\ \rho(A)<1$ and the sets $\drivers,\ \outputs$.
	For the edge modification $\{(\source,\target),w\}$, $-A_{\target \source}\leq w < 1/(\left(I - A\right)^{-1})_{\source\target }$, it holds
	\begin{align}
	||G^\delta||_\hinf =
	\frac{\sqrt{\sum\limits_{o \in \outputs} \left(\left( (I  - A ^{-1}\right)_{o \target}\right) ^2}\ |w| \ \sqrt{\sum\limits_{k \in \drivers} \left(\left( (I  - A )^{-1}\right)_{\source k}\right) ^2}}{1 - \left((I- A)^{-1}\right)_{\source \target }w }
	\label{eq:hinfGainGDelta}
	\end{align}
	\label{cor:HinfSingleEdge}		
\end{theorem}
\proof
	The condition $-A_{\target \source}\leq w < 1/(\left(I - A\right)^{-1})_{\source \target}$ implies that $\bar{A} \geq 0$, $\rho(\bar{A})< 1$ and the norm $||G^\delta||_\hinf$ is bounded. 
	
	For positive weights, $0 < w < 1/(\left(I - A\right)^{-1})_{\source \target}$, $G^\delta$ is externally positive, hence $||G^\delta||_\htwo = \bar{\sigma}(G^\delta(0))$.
	Notice that $G^\delta(0)$ is a rank one matrix. We can write
	\begin{align}
		G^\delta(0) &= \tf{A}{\outputs \target}(0) G^\delta_c(0) \tf{A}{\source \drivers}(0) \nonumber \\
		&= \frac{\tf{A}{\outputs \target}(0)\sign(G^\delta_c(0))}{|\tf{A}{\outputs \target}(0)|}  \nonumber \\ 
		&\quad \ \ \cdot   |\tf{A}{\outputs \target}(0)|\ |G^\delta_c(0)|\ |\tf{A}{\source \drivers}(0)| \ \frac{\tf{A}{\source \drivers}(0)}{|\tf{A}{\source \drivers}(0)|},
		\label{eq:GdeltaThetaFreq}
	\end{align}
	where $\sign(\cdot)$ is the signum function.
	The equation \eqref{eq:GdeltaThetaFreq} is a singular value decomposition since 
	$\tf{A}{\source \drivers}(0)/|\tf{A}{\source \drivers}(0)|$
	and $\tf{A}{\outputs \target}(0)\sign(G^\delta_c(0))/|\tf{A}{\outputs \target}(0)|$ are norm-1 row resp. column vectors. 
	By identification, the positive scalar \break $|\tf{A}{\outputs \target}(0)|\ |G^\delta_c(0)| \ |\tf{A}{\source \drivers}(0)|$ is the only (hence the maximal) singular value.
	Evaluating it gives \eqref{eq:hinfGainGDelta}.
	
	For $-A_{\target \source} \leq w < 0$, instead $-G^\delta$ is externally positive. In this case, replace $G^\delta_c(0)$ with $-G^\delta_c(0)$ in equation \eqref{eq:GdeltaThetaFreq} to obtain a singular value decomposition of $-G^\delta(0)$. Then use $||G^\delta||_\hinf = ||-G^\delta||_\hinf = \bar{\sigma}(-G^\delta(0))$.~ 
\qed

The computational complexity in evaluating the stability bounds of Theorem \ref{prop:stabRad} and $||G^\delta||_\hinf$ lies in the matrix inversion. 
This however has to be done only once for all possible edge modifications $\{(\source,\target),w\}$. 

\subsection{The $\htwo$ norm of $G^\delta$}

The next lemma establishes two properties of positive systems that will be used to bound $||G^\delta||_\htwo$.
\begin{lemma}
	Let $G$ and $H$ be two externally positive systems with impulse responses $g(t)$ resp. $h(t)$. Assuming matching input/output dimensions, the following hold,
	\begin{itemize}
		\item[P1.] $||G + H||_\htwo^2 \geq ||G||^2_\htwo + ||H||^2_\htwo$. \label{prop:prypertyParallel}
		\item[P2.] For $G$ a multiple input single output (MISO) system and $H$ a single input multiple output (SIMO) system, $||H G||_\htwo \geq ||H||_\htwo  ||G||_\htwo$. \label{prop:prypertySeries}
	\end{itemize}
	\label{prop:posProperties}
\end{lemma}
\proof
	
	P1: From the definition of the $\htwo$ norm \eqref{eq:H2normGeneral} we have
	\begin{align*}
		||G + H||_\htwo^2 	&= \trace \left( \sum_{t=0}^\infty (g(t) + h(t)) (g(t) + h(t))^\top \right) \\
		&\geq \trace \left( \sum_{t=0}^\infty g(t)g(t)^\top \right) + \trace \left( \sum_{t=0}^\infty
		h(t)h(t)^\top \right)	\\
		&= ||G||_\htwo^2 + ||H||_\htwo^2.
	\end{align*}
	In the inequality we disregard the cross-terms, knowing that they are non-negative since $g(t) \geq 0$ and $h(t) \geq 0\ \forall t$.
	
	P2: We first show the result for $G$ and $H$ being SISO systems. The impulse response of $H G$ is the convolution $(h * g)(t)$, hence
	\begin{align*}
	||HG||_\htwo^2 
	&= \sum_{t=0}^{\infty} (h * g)^2(t) 
	= \sum_{t=0}^{\infty} \left( \sum_{\tau = 0}^t g(t-\tau)h(\tau)\right)^2 \\
	&\geq \sum_{t=0}^{\infty} \sum_{\tau = 0}^t g^2(t-\tau) h^2(\tau) 
	= \sum_{\tau = 0}^\infty \sum_{t = \tau}^\infty g^2(t-\tau) h^2(\tau)  \\
	&= /\text{using } k = t-\tau / = \sum_{\tau = 0}^\infty \sum_{k = 0}^\infty g^2(k) h^2(\tau) \\
	&= \sum_{\tau = 0}^\infty  h^2(\tau) \sum_{k = 0}^\infty g^2(k) = ||H||_\htwo^2 ||G||_\htwo^2.
	\end{align*}
	In the inequality above we use the fact that all cross terms are non-negative.
	
	Next, let $G = [G_{1}\ \dots \ G_{n_{\drivers}}]$ be an $n_\drivers$-input-single-output system, and $H = [H_{1}\ \dots \ H_{n_\outputs}]^\top$ a single-input-$n_\outputs$-output system. Then $HG$ is $n_\outputs\times n_\drivers$, and the transfer function from the $i$-th input to the $j$-th output is $(HG)_{ji} = H_{j}G_{i}$. From \eqref{eq:H2normFromGramian_1a},
	\begin{align*}
	&||HG||_\htwo^2 = \sum_{j=1}^{n_\outputs} \sum_{i=1}^{n_\drivers} ||(HG)_{ji}||_\htwo^2 = \sum_{j=1}^{n_\outputs} \sum_{i=1}^{n_\drivers} ||H_{j} G_{i}||_\htwo^2\\
	&\geq \sum_{j=1}^{n_\outputs} \sum_{i=1}^{n_\drivers} ||H_{j}||_\htwo^2 ||G_{i}||_\htwo^2 = \sum_{j=1}^{n_\outputs} ||H_{j}||_\htwo^2 \sum_{i=1}^{n_\drivers} ||G_{i}||_\htwo^2 \\
	&= ||H||_\htwo^2 ||G||_\htwo^2,
	\end{align*}
	which proves the lemma. In the inequality we use P2 for SISO systems.  
\qed

\begin{remark}
	Notice that P2 does not hold in general for $G$ and $H$ positive multiple-input-multiple-output (MIMO) systems. For instance, with $G=H$ both given by $[y_1(t)\ y_2(t)]^\top = [u_1(t-1)\ u_2(t-1)]^\top$, it is $||HG||^2_\htwo = 2 < ||H||^2_\htwo ||G||^2_\htwo = 4$. 
	Moreover, the properties P1 and P2 do not hold in general for $G,\ H$ \emph{not} positive.
\end{remark}

\begin{theorem}
	Consider a network with adjacency matrix $A\geq 0,\ \rho(A)<1$ and the sets $\drivers,\ \outputs$.
	For the edge modification $\{(\source,\target),w\}$, $-A_{\target \source}\leq w < 1/(\left(I - A\right)^{-1})_{\source \target }$, the $\htwo$-norm of the delta system is bounded by
	\begin{align}
	||G^\delta||_\htwo^2 \geq  p_\target \ \frac{  w^2}{1- \varepsilon_{\target\to \source}w^2 }\ q_\source. \label{eq:positiveLowerBound}
	\end{align}
	\label{thm:H2LowerBound}
\end{theorem}
\proof
	The conditions
	\begin{align*}
	&
	\begin{cases}
		1 > w(\left(I - A\right)^{-1})_{\source \target } = w \left(I + A + A^2 + \dots \right)_{\source \target},\\
		w > 0,\\
		 A\geq 0 
	\end{cases}
	\end{align*}
	imply
	\begin{align*}
	& 1 > w^2 \left(\left(I + A + A^2 + \dots \right)_{\source \target}\right)^2 \\
	& \ \ \geq w^2 \left( (I_{\source \target})^2 + (A_{\source \target})^2 + ((A^2)_{\source \target})^2 + \dots \right) \\ 
	& \ \ = w^2 \varepsilon_{\target \to \source}.
	\end{align*}
	It follows from Definition \ref{def:posSystem} that externally positive systems in series or in parallel constitute an externally positive system. Hence we conclude that the feedback loop 
	\begin{align*}
		G_c^\delta = w\left( 1 + \tf{A}{\source \target}w + (\tf{A}{\source \target}w)(\tf{A}{\source \target}w) +\dots  \right)
	\end{align*}
	is externally positive, and 
	\begin{align*}
	&||G^\delta_c||_\htwo^2
	= \left|\left|w\left( 1 + \tf{A}{\source \target}w + (\tf{A}{\source \target}w)(\tf{A}{\source \target}w) +\dots  \right) \right|\right|^2_\htwo
	\\ &\geq w^2 \left( 1 +  ||\tf{A}{\source \target}w||_\htwo^2 + ||\tf{A}{\source \target}w\tf{A}{\source \target}w||_\htwo^2 + \dots \right) 
	&
	\\ &\geq w^2 \left( 1 +  ||\tf{A}{\source \target}w||_\htwo^2 + ||\tf{A}{\source \target}w||_\htwo^2 ||\tf{A}{\source \target}w||_\htwo^2 + \dots \right) 
	\\ & = \frac{w^2}{1 - \varepsilon_{\target \to \source}w^2},
	\end{align*}
	where the properties P1 and P2 are used in the first resp. second inequality. We also use $||\tf{A}{\source \target}w||^2_\htwo = \varepsilon_{\target \to \source}w^2 < 1$ and geometric series. Finally, since $G^\delta = \tf{A}{\outputs \target} G_c^\delta \tf{A}{\source \drivers}$, i.e. three positive systems in series, we can apply Property P2 to obtain
	\begin{align*}
	||G^\delta ||_\htwo^2 &\geq
	\left|\left|\tf{A}{\outputs \target}\right|\right|_\htwo^2 
	\left|\left|G^\delta_c\right|\right|_\htwo^2
	\left|\left|\tf{A}{\source \drivers}\right|\right|_\htwo^2 \\
	&= 
	\frac{p_\target w^2 q_\source}{1- \varepsilon_{\target\to \source}w^2 }.
	\end{align*}
	 
\qed

\subsection{Interpretations}

The impact of an edge modification in networks with stable dynamics is quantified with the $\hinf$ norm of the delta system in Theorem \ref{cor:HinfSingleEdge} and with the $\htwo$ norm in Theorem \ref{thm:H2LowerBound}. 
Although the mathematical expressions differ between the two cases, we can see that the impact depends on three things (besides the actual weight $w$):
\begin{enumerate}[label=(\roman*)]
	\item The gain from the input nodes to the source node of the modified edge:
	\begin{itemize}
		\item[] $\hinf$: $|\tf{A}{\source \drivers}(0)|$,
		\item[] $\htwo$: the input-to-node centrality $q_\source$.
	\end{itemize}
	
	\item The gain from the target node to the output nodes:
	\begin{itemize}
		\item[] $\hinf$: $|\tf{A}{\outputs \target}(0)|$,
		\item[] $\htwo$: the node-to-output centrality $p_\target$.
	\end{itemize}
	
	\item The strength of the feedback loop: 
	\begin{itemize}
		\item[] $\hinf$: $\tf{A}{\source \target}(0) = \left((I-A)^{-1}\right)_{\source \target}$,
		\item[] $\htwo$: the walk energy $\varepsilon_{\target \to \source}$.
	\end{itemize}
	
\end{enumerate}
These dependencies are evident in the block diagram of Figure~\ref{fig:deltaSystem}.


\section{Edge addition in undirected networks with Laplacian dynamics}
\label{sec:LaplacianDynamics}

In this section we consider undirected connected consensus networks obtained as discretizations of  Laplacian dynamics. Let $\mathcal{K} = \{k_1,\dots,k_{n_\mathcal{K}} \} \subseteq \mathcal{V}$ be the input nodes and define $f_{i}(t) = u_\ell(t)$ if $i=k_\ell \in \mathcal{K}$ and $f_{i}(t) = 0$ otherwise. The state of node $i \in \mathcal{V}$ is updated in discrete time according to
\begin{align*}
x_i(t+1) = x_i(t) + \sum_{j \text{ s.t. } (j,i)\in \mathcal{E}} w_{ij}(x_j(t) - x_i(t)) + f_i(t)
\end{align*}
with $ \mathcal{A}=[w_{ij}]$ an irreducible adjacency matrix.
As in the previous section, we assume that the edge weights are always positive.
The graph Laplacian matrix 
\begin{align}
L = \sum_{(i,j) \in \mathcal{E}} w_{ij}e_{ij}e_{ij}^\top \label{eq:undirGraphLapFromEdges}
\end{align}
is positive semidefinite, with $\onevec$ both a left and right eigenvector corresponding to the zero eigenvalue.
With $L$, the dynamics of the entire network can be expressed as
\begin{align}
	x(t+1)  &= (I-L) x(t) + Bu(t), \label{eq:ConsensusNw} 
\end{align}
which is on the form \eqref{eq:discreteStateUpdate} with the state update matrix $A = I - L$ and input matrix $B = E_\mathcal{K}$.
Apart from connectivity of the network, we further assume that 
\beq
\max_i(\lambda_i(L)) =\rho(L) <1 
 \label{eq:stepsize}
 \eeq
 where $ \lambda_i(L) $ is an eigenvalue of $L$. 
 Since $L$ is symmetric and positive semidefinite, from Schur theorem (see \cite{horn2012matrix}, Thm.~4.3.45) \eqref{eq:stepsize} implies that for each $i \in \mathcal{V}$ it holds
\begin{align*}
	\sum_{j \text{ s.t. } (j,i) \in \mathcal{E}} w_{ij} < 1. 
\end{align*}
If the model \eqref{eq:ConsensusNw} is obtained as a discretization of a continuous time model, then the condition \eqref{eq:stepsize} can always be satisfied by choosing the step-size small enough.
Under these assumptions, $ A \geq 0 $ and its eigenvalues, $ \lambda_i$, $i=1, \ldots, n$, are positive real and can be ordered as $0 < \lambda_1 \leq \lambda_2 \leq \dots \leq \lambda_{n-1} < \lambda_n = 1$,
where $\onevec$ now is the left and right eigenvector corresponding to $\lambda_n$, i.e., $ A$ is a row and column stochastic matrix \cite{horn2012matrix}. Hence, the dynamical system \eqref{eq:ConsensusNw} is marginally stable and its infinite time controllability Gramian is unbounded. 
Consequently, if we consider all nodes as output nodes (i.e. $C=I$), then the $\htwo$ norm of the system is unbounded. 
A similar problem occurs for the $ \hinf $ norm.
Adding an edge to the network does not change this fact: the sum \eqref{eq:undirGraphLapFromEdges} is extended with one more term but  $(I-L)\onevec = \onevec$ still holds, which means $\lambda_n = 1$, i.e., the system remains marginally stable.

However, a small modification of \eqref{eq:ConsensusNw} gives bounded $\htwo$/$\hinf$ norms and meaningful interpretations of the effects of edge additions.
Let 
\begin{align*}
	J &= \lim_{t \to \infty} A^t = \onevec \onevec^\top/n.
\end{align*}
The following properties of $J$ can be computed straightforwardly.
\begin{lemma}
	For the matrix $J$ it holds:
	\begin{align*}
		&J^k = J,\ \forall k \in \mathbb{N} \Rightarrow (I-J)^k = (I-J)\ \forall k \in \mathbb{N}, \\
		&AJ = J, \\ 
		&A(I-J) = (I-J)A.
	\end{align*}
	\label{lem:J}
\end{lemma}
Define the \emph{displacement system} with state vector $\xi(t) = (I-J)x(t)$,
\begin{align}
	 \xi(t+1) &= (I-J) \left( A x(t) + B u(t)\right) \nonumber \\
	&= A_J \xi(t) + B_J u(t), \label{eq:displacementSystem}
\end{align}
where $A_J = (I-J)A$ and $B_J = (I-J)B$. Note that since $I - J$ is singular, $\xi = (I-J)x$ is not a proper change of basis, rather a projection onto the subspace orthogonal to $\text{span}(\onevec)$, the \emph{disagreement subspace} \cite{siami2013fundamental,young2010robustness}.
\begin{proposition}
\label{prop:displamente-stability}
	The displacement system \eqref{eq:displacementSystem} is internally stable.
\end{proposition}
\proof
	Since $A$ is symmetric, it can be factorized 
	\begin{align*}
	&A = U \Lambda U^\top,
	\end{align*}
	where $U\in \mathbb{R}^{n\times n}$ is orthogonal ($UU^\top = U^\top U = I$), $\Lambda = \diag(\lambda_{1}, \dots, \lambda_{n})$, and the $n$-th column of $ U$, $U[n] = \onevec / \sqrt{n}$, is the normalized eigenvector corresponding to $\lambda_n = 1$.	
	For $A_J$ we have
	\begin{align} 
	A_J &= U \Lambda U^\top (I - \onevec \onevec^\top / n) \nonumber \\
	&= \left(\sum_{i=1}^n U[i]\lambda_i U[i]^\top \right) (I - U[n]U[n]^\top) \nonumber \\
	&= \sum_{i=1}^n U[i]\lambda_i U[i]^\top - U[n]\lambda_n \underbrace{U[n]^\top U[n]}_{=1} U[n]^\top \nonumber \\
	&= \sum_{i=1}^{n-1} U[i]\lambda_i U[i]^\top = U\tilde{\Lambda} U^\top, \label{eq:factorizationAJ}
	\end{align}
	where $\tilde{\Lambda} = \diag (\lambda_{1},\dots,\lambda_{n-1},0)$. Hence the spectral radius is $\rho(A_J) = \lambda_{n-1} < 1$.
\qed

Since the displacement system is internally stable, its infinite time controllability Gramian
\begin{align}
W  _\xi  &=\sum_{\tau=0}^\infty A_J^\tau B_J B_J^\top (A_J^\top)^\tau \label{eq:scherpenP123}
\end{align}
exists and is the unique solution to the Lyapunov equation
\begin{align*}
	A_J W  _\xi  A_J^\top -W  _\xi  + B_J B_J^\top = 0.
\end{align*}
Using the properties of $J$ stated in Lemma~\ref{lem:J}, this equation can alternatively be expressed as
\begin{align*}
\begin{cases}
A W  _\xi  A^\top -W  _\xi  + B_J B_J^\top ,\\
J W  _\xi  J^\top = 0.
\end{cases}
\end{align*}
The continuous time equivalent to the Gramian \eqref{eq:scherpenP123} is referred to as the \emph{pseudo} controllability Gramian of \eqref{eq:ConsensusNw} in \cite{cheng2019Gramian}.

Notice that $\text{span}(\onevec)$ is in the nullspace of $W  _\xi $. With $W  _\xi $, the energy needed to control the network in any direction of the $(n-1)$-dimensional disagreement subspace can be computed.


\subsection{Transfer function for the displacement system and its $ \hinf$ norm}

\begin{proposition}
\label{prop:sameTF}
The two systems

\[
\Upsilon_\xi  = \{ A_J, \, B_J, \, C \} \quad \text{and} \quad \Upsilon_\chi = \{ A, \, B_J, \, C \} 
\]
correspond to nonminimal realizations of the same transfer function, but with different zero/pole cancellations.
Furthermore, the $ \htwo  $ and $ \hinf $ norms of $ \Upsilon_\xi  $ and $ \Upsilon_\chi $ are the same.
\end{proposition}
\proof
From Lemma~\ref{lem:J}, for the impulse responses we have:
\beq
g_x(t) = C A^{t-1}B_J = C A^{t-1} (I-J) B= C A_J^{t-1} B_J = g_\xi (t).
\label{eq:impulse_resp_lapl}
\eeq
If the impulse responses are identical, so are the corresponding transfer functions, modulo zero/pole cancellations. 
Since $A$ and $A_J $ differ only for the eigenvalue $ \lambda_n=1$, in $ \Upsilon_\chi $ the term $ z -1 $ corresponds to the zero/pole cancellation, while in $ \Upsilon _\xi  $ it is the term $ z$. 
Since the two transfer functions are the same, so are the corresponding $ \htwo  $ and $ \hinf $ norms. 
For $ \htwo  $ it is obvious from \eqref{eq:impulse_resp_lapl} and \eqref{eq:H2normGeneral}, while for $ \hinf$ it will be obvious once we have proven  Proposition~\ref{prop:DC-gain-lapl}.
\qed

\begin{remark}
By the same token as in Proposition~\ref{prop:sameTF}, also the systems $ \{ A, \, B, \, C_J \} $, $ \{ A, \, B_J, \, C_J \} $, $ \{ A_J, \, B, \, C \} $, $ \{ A_J, \, B, \, C \} $, $ \{ A_J, \, B, \, C_J \} $, $ \{ A_J, \, B_J, \, C_J \} $ all have the same transfer function as $ \Upsilon_\chi$, $ \Upsilon_\xi $, modulo zero/pole cancellations.
\end{remark}

The following proposition states that although the displacement system is not externally positive, its $ \hinf $ norm is still given by the spectral norm of its DC-gain.
\begin{proposition}
\label{prop:DC-gain-lapl}
For the displacement system $ \Upsilon_\xi $ it is $ || G _\xi||_\hinf = \bar{\sigma}(G _\xi (0)) $, where $ G _\xi (z) $ is the transfer function of $ \Upsilon_\xi  $. 
\end{proposition}
\proof
From \eqref{eq:impulse_resp_lapl}, Lemma~\ref{lem:J}, and \eqref{eq:factorizationAJ}, one gets
\[
\begin{split}
G _\xi (z) & =  \sum_{\tau=0}^\infty C A_J^\tau B_J z^{-\tau} =  \sum_{\tau=0}^\infty C A_J^\tau  B z^{-\tau} 
\\
& =  \sum_{\tau=0}^\infty C \left( U\tilde{\Lambda} U^\top\right)^\tau B z^{-\tau}  = C U \left( \sum_{\tau=0}^\infty \tilde{\Lambda}^\tau  z^{-\tau} \right) U^\top  B \\
& =   C  U \begin{bmatrix} \sum_{\tau=0}^\infty \lambda_1^\tau z^{-\tau} & \\ & \ddots \\ & & \sum_{\tau=0}^\infty \lambda_{n-1} ^\tau z^{-\tau} \\ & & & 0 \end{bmatrix} U^\top B\\
& =   C U \begin{bmatrix} \frac{1}{z-\lambda_1} & \\ & \ddots \\ & & \frac{1}{z -\lambda_{n-1} }\\ & & & 0 \end{bmatrix} U^\top B \\
\end{split}
\]
Since $ \lambda_i > 0$, $i=1,\ldots, n-1 $, $ \frac{1}{z-\lambda_i } $ is the transfer function of a positive system, hence $ {\rm argsup}_\theta \sigma \left( \frac{1}{e^{i \theta } -\lambda_i} \right) =0$, $ i=1, \ldots, n-1$. Furthermore, since $ U$ is an isometry, and $B, \, C $ are (nonnegative) elementary columns/rows matrices, it follows that $ \sup_\theta \bar{\sigma} \left(G _\xi (\theta) \right) = \bar{\sigma}(G _\xi (0)) $.
\qed


\subsection{An upper bound on the $ \hinf$ norm for edge additions}

From \eqref{eq:undirGraphLapFromEdges}, the edge addition $\{(\source, \target), w\}$, $ w>0$, results in the Laplacian matrix
\[
\bar{L} = L + w e_{\source \target} e_{\source \target}^\top \\
\]
and hence in
\beq
\bar{A} = A - w e_{\source \target} e_{\source \target}^\top ,
\label{eq:Lapl-A-addition}
\eeq
where we always assume that the weights of $ \bar A $ satisfy \eqref{eq:stepsize}.
Since we are dealing with an undirected graph, \eqref{eq:Lapl-A-addition} is always intended as  the edge pair $ \{ (\source, \target), w\} $ and $ \{ (\target, \source), w\}$, and leads to 4 terms, because two diagonal terms are needed to guaranteee that the Laplacian structure of $\bar L$ (and the stochasticity of $ \bar A$) is preserved. 
The expression \eqref{eq:Lapl-A-addition} leads to a similar expression for the displacement system
\[
\bar{A}_J = \bar{A} (I - J) = A_J - w e_{\source \target} e_{\source \target}^\top.
\]
This can be used to compute the delta system for the displacement
\beq
G _\xi ^\delta = \tf{\bar{A}_J} {\outputs \drivers}-  \tf{A_J} {\outputs \drivers} .
\label{eq:G_xi_delta}
\eeq
By construction, $ \tf{A}{\outputs \drivers} $ and $ \tf{\bar{A}}{\outputs \drivers} $ are both marginally stable, while $ \tf{A_J}{\outputs \drivers} $ and $ \tf{\bar{A}_J}{\outputs \drivers} $ are both stable, since both are displacement systems and Proposition~\ref{prop:displamente-stability} applies. 
From \eqref{eq:G_xi_delta}, then, also $ G_\xi^\delta $ is stable (``parallel interconnection'').
\begin{proposition}
It holds
\beq
\begin{split}
 G _\xi ^\delta = & \left( \tf{A_J}{\outputs \source} - \tf{A_J}{\outputs \target} \right)  \\
& \cdot \left( 1 - w \left( \tf{A_J}{\source \target} + \tf{A_J}{\target \source} - \tf{A_J}{\source \source} - \tf{A_J}{\target \target} \right) \right)^{-1} w \\ 
& \cdot  \left( \tf{A_J}{ \target \drivers} - \tf{A_J}{ \source \drivers} \right) .
\end{split}
\label{eq:Gz-delta}
\eeq
\end{proposition}
\proof
The proof is similar to that of Proposition~\ref{prop:deltaSystem}, with the extra difficulty of having to deal with 4 terms simultaneously. 
Repeating the calculations in the proof of Proposition~\ref{prop:deltaSystem}, we get
\[
G _\xi ^\delta =  \left( \tf{\bar{A}_J}{\outputs \source} - \tf{\bar{A}_J}{\outputs \target} \right) w  \left( \tf{A_J}{ \target \drivers} - \tf{A_J}{ \source \drivers} \right) ,
\]
and then
\begin{align*}
\tf{\bar{A}_J}{\outputs \source} - \tf{A_J}{\outputs \source} & = \left( \tf{\bar{A}_J}{\outputs \source} - \tf{\bar{A}_J}{\outputs \target} \right) w \left( \tf{A_J}{ \target \source} - \tf{A_J}{ \source \source} \right) \\
\tf{\bar{A}_J}{\outputs \target} - \tf{A_J}{\outputs \target} & = \left( \tf{\bar{A}_J}{\outputs \target} - \tf{\bar{A}_J}{\outputs \source} \right) w \left( \tf{A_J}{ \source \target } - \tf{A_J}{ \target  \target} \right) .
\end{align*}
Hence
\[
\begin{split}
& \tf{\bar{A}_J}{\outputs \source} - \tf{\bar{A}_J}{\outputs \target} =   \left(  \tf{A_J}{\outputs \source} - \tf{A_J}{\outputs \target} \right) \\
& \qquad \cdot \left( 1 -w \left( \tf{A_J}{\source \target} + \tf{A_J}{\target \source} - \tf{A_J}{\source \source} - \tf{A_J}{\target \target} \right) \right)^{-1} .
\end{split}
\]
from which the result follows.
\qed

Notice that the middle part of $ G _\xi ^\delta $,
\[
\begin{split}
G_c^\delta = &  \left( 1 -w \left( \tf{A_J}{\source \target} + \tf{A_J}{\target \source} - \tf{A_J}{\source \source} - \tf{A_J}{\target \target} \right) \right)^{-1} w \\
= & \left( 1 + w \left[\begin{array}{c|c}
	A_J & e_{\source \target} \\ \hline e_{\source \target}^\top & 0 \Tstrut
	\end{array}\right] \right)^{-1} w
	\end{split}
\]
has still the interpretation of all additional feedback loops induced by the new edge(s), including self-loops. 

Unlike in Section~\ref{sec:stableDynamics}, we can only provide an upper bound on the $ \hinf$ norm of $ G _\xi ^\delta $, not an exact value. 

\begin{theorem}
Consider a network with graph Laplacian $L$ and sets $ \drivers$, $ \outputs$. 
Consider an edge addition $ \{ (\source, \target), w\}$, $ w>0$, which respects \eqref{eq:stepsize}.
For the $ \delta $ system of the corresponding displacement system, it is
\beq
|| G _\xi ^\delta ||_\hinf \leq \frac{ \left( 
\gamma_{\outputs \source} \gamma_{\target \drivers} 
+  \gamma_{\outputs \target } \gamma_{\source \drivers} 
+ \gamma_{\outputs \target } \gamma_{\target \drivers}  
+\gamma_{\outputs \source} \gamma_{\source \drivers} 
\right) w}{\left| 1 - w ( L_{\source \target}^\dagger +  L_{ \target \source}^\dagger -  L_{\source \source}^\dagger -  L_{\target \target}^\dagger  ) \right|} 
\label{eq:hing_delta_displ}
\eeq
where, for $ r=\source, \target$, 
\[
\gamma_{\outputs r}  =  \sqrt{ \sum_{o\in \outputs}  \left( L_{o r}^\dagger +\frac{1}{n} \right)^2 }, \quad 
\gamma_{r \drivers}  =  \sqrt{ \sum_{k\in \drivers}  \left( L_{ r k}^\dagger +\frac{1}{n} \right)^2 },
\] 
and $ L_{r v}^\dagger $  is the $ (r, v) $ entry of the pseudoinverse of the Laplacian $L$. 
\end{theorem}
\proof
The proof follows the same reasoning as the one of Theorem~\ref{cor:HinfSingleEdge}.
From Proposition~\ref{prop:DC-gain-lapl},  $||G _\xi ^\delta||_\hinf = \bar{\sigma}(G _\xi ^\delta(0))$.
Computing the DC gain of $ G _\xi ^\delta$  from \eqref{eq:Gz-delta}, 
\begin{align*}
G _\xi ^\delta(0) &= \frac{\tf{A_J}{\outputs \source}(0)\sign(G^\delta_c(0))}{|\tf{A_J}{\outputs \source}(0)|} \ \nonumber \\ 
&  \quad \ \ \cdot  |\tf{A_J}{\outputs \source}(0)|\ |G^\delta_c(0)|\ |\tf{A_J}{\target \drivers}(0)| \ \frac{\tf{A_J}{\target \drivers}(0)}{|\tf{A_J}{\target \drivers}(0)|} \\
& +\frac{\tf{A_J}{\outputs \target}(0)\sign(G^\delta_c(0))}{|\tf{A_J}{\outputs \target}(0)|} \nonumber \\ 
&\quad \  \ \cdot  |\tf{A_J}{\outputs \target}(0)|\ |G^\delta_c(0)|\ |\tf{A_J}{\source \drivers}(0)| \ \frac{\tf{A_J}{\source \drivers}(0)}{|\tf{A_J}{\source \drivers}(0)|} \\
& -\frac{\tf{A_J}{\outputs \target}(0)\sign(G^\delta_c(0))}{|\tf{A_J}{\outputs \target}(0)|} \nonumber \\ 
&\quad \ \ \cdot   |\tf{A_J}{\outputs \target}(0)|\ |G^\delta_c(0)|\ |\tf{A_J}{\target \drivers}(0)| \ \frac{\tf{A_J}{\target \drivers}(0)}{|\tf{A_J}{\target \drivers}(0)|} \\
& - \frac{\tf{A_J}{\outputs \source}(0)\sign(G^\delta_c(0))}{|\tf{A_J}{\outputs \source}(0)|}\nonumber \\ 
&\quad \  \ \cdot   |\tf{A_J}{\outputs \source}(0)|\ |G^\delta_c(0)| \ | \tf{A_J}{\source \drivers}(0)| \ \frac{\tf{A_J}{\source \drivers}(0)}{|\tf{A_J}{\source \drivers}(0)|} .
\end{align*}
Each of the 4 terms above is a rank-1 matrix which can be considered in singular value decomposition form, with the inner term ($  |\tf{A_J}{\outputs \source}(0)|\ |G^\delta_c(0)|\ |\tf{A_J}{\target \drivers}(0)|  $, and so on for the others) being the corresponding singular value. 
Since $ I - A_J = L+J$ and $ (L+J)^{-1} = L^\dagger + J$ \cite{Xiao2003Resistance}, it is $ (I-A_J)^{-1} = L^\dagger + J$, hence when computing the Euclidean norm of the various terms, we have
\[
\begin{split}
\left| \tf{A_J}{\outputs \source}(0) \right| =  &\sqrt{ \sum_{o\in \outputs} \left( ( (I - A_J )^{-1} )_{o \source} \right)^2 } = 
\sqrt{ \sum_{o\in \outputs} \left( (L^\dagger +J )_{o \source} \right)^2 } \\
= &  
\sqrt{ \sum_{o\in \outputs} \left( L^\dagger_{o\source} +\frac{1}{n} \right)^2 }
\end{split}
\]
and similarly for the other transfer functions. 
Using the same argument for $ G_c^\delta $, it is
\[
G_c^\delta(0) = \frac{w}{\left| 1 - w ( L_{\source \target}^\dagger + \frac{1}{n} +  L_{ \target \source}^\dagger + \frac{1}{n} -  L_{\source \source}^\dagger - \frac{1}{n} -  L_{\target \target}^\dagger  - \frac{1}{n}  ) \right|}
\]
In the worst case possible, the 4 terms of the singular value decompositions will be collinear, with the overall singular value given by the sum of the 4 terms, hence in general  in \eqref{eq:hing_delta_displ}  we have an inequality sign.
\qed

\begin{remark}
The denominator of \eqref{eq:hing_delta_displ} can be expressed compactly in terms of the effective resistance between nodes $ \source $ and $ \target $ \cite{Xiao2003Resistance}:
\[
R_{\target \source } = e_{\target \source }^\top L^\dagger   e_{\target \source } =  
L_{\source \source}^\dagger +  L_{\target \target}^\dagger  -L_{\source \target}^\dagger -  L_{ \target \source}^\dagger .
\]
\end{remark}


\subsection{Coherence as $\htwo$ norm}
In order to compute $ \htwo$ norms, we will  consider only the case of inputs and outputs on each node.
The following result is similar to \cite{LOVISARI2012Performance}, but it is here reformulated for the displacement system.
\begin{theorem}
	In the displacement system \eqref{eq:displacementSystem}, let $0 < \lambda_1 \leq \lambda_2 \leq \dots \leq \lambda_{n-1} < \lambda_n = 1$ be the eigenvalues of $A$, and $B=C=I$. Then the $\htwo$ norm is given by 
	\begin{subequations}
		\begin{align}
			||\Upsilon_\xi ||_\htwo^2 &= \sum_{i = 1}^{n-1} \frac{1}{1 - \lambda_i^2} \\
			&=  \trace ((I-A^2)^\dagger). \label{eq:coherenceDiscreteEigs}
		\end{align}
	\end{subequations}

\end{theorem}
\proof
From $B=I $, it is $ B_J = I-J$, and it is easy to show (using Lemma \ref{lem:J} and the factorization \eqref{eq:factorizationAJ}) that 
	\begin{align*}
		A_J^\tau B_J &= \sum_{i=1}^{n-1} U[i]\lambda_i^\tau U[i]^\top,\ \forall \tau \in \mathbb{N}_0 \\
		\Rightarrow	
		W _\xi  &= \sum_{\tau=0}^\infty A_J^\tau B_J B_J^\top (A_J^\top)^\tau = \sum_{\tau = 0}^\infty \sum_{i=1}^{n-1} U[i] (\lambda_i^2)^\tau U[i]^\top
	\end{align*}
	and so
	\begin{align*}
	||\Upsilon_\xi ||_\htwo^2 &= \trace(W _\xi ) 
	= \sum_{\tau = 0}^\infty \sum_{i=1}^{n-1} (\lambda_i^2)^\tau = \sum_{i = 1}^{n-1} \frac{1}{1 - \lambda_i^2}.
	\end{align*}
	Moreover, since
	\begin{align*}
	I-A^2 = UU^\top -U\Lambda U^\top U \Lambda U^\top = U(I-\Lambda^2)U^\top,
	\end{align*}
	where $I- \Lambda^2$ is diagonal, the eigenvalues of $I - A^2$ are the diagonal elements ${1-\lambda_i^2}$, $i=1,\dots,n$. 
	The matrix $I-A^2$ is singular since $1-\lambda_n^2 = 0$ is one eigenvalue.
	It is easy to verify that the matrix 
	\begin{align}
	(I-A^2)^\dagger = U \diag \left(\frac{1}{1-\lambda_{1}^2}, \dots, \frac{1}{1-\lambda_{n-1}^2},0 \right) U^\top \label{eq:ImAsqPinv}
	\end{align}
	is the pseudo-inverse of $I-A^2$. Finally, since the trace of a matrix is the sum of its eigenvalues, we conclude that 
	\begin{align*}
	||\Upsilon_\xi ||_\htwo^2 =  \trace ((I-A^2)^\dagger).
	\end{align*}
\qed

Notice that in terms of $ A_J $ the $ \htwo$ norm can be expressed as
\[
||\Upsilon_\xi ||_\htwo^2 = \trace ((I-A_J^2)^{-1}) -1.
\]

When the inputs to the network are noise processes rather than control signals, then the infinite time controllability Gramian can be interpreted as the covariance matrix of the state vector \cite{rugh1996linear}. The network model 
\begin{align}
x(t+1) &= (I- L) x(t) + w(t),
\label{eq:coherenceNw}
\end{align}
where $w(t)$ is a vector of zero-mean i.i.d. white noise processes, is studied in for instance \cite{bamieh2012coherence,hagberg2008rewiring,siami2017growing}. This model assumes noise inputs on each node. One performance metric for such networks is the network \emph{coherence}, usually defined as the steady state variance of the deviation from consensus \cite{bamieh2012coherence}, in equations,
\begin{align}
\mathcal{C} &= \lim_{t \to \infty} \sum_{i = 1}^n \mathbb{E} \left[\left( x_i(t) - \frac{1}{n} \sum_{j=1}^n x_j(t) \right)^2 \right]. \label{eq:coherenceTrace}
\end{align}
For the network model \eqref{eq:coherenceNw}, this expression can be developed as
\begin{align*}
\mathcal{C} &= \lim_{t \to \infty} \mathbb{E} \left[ \sum_{\tau=0}^{t-1} \left(A_J^{(t-\tau)}B_J w(\tau) \right)^\top A_J^{(t-\tau)}B_J w(\tau) \right] \\
&= \trace (W _\xi )
\end{align*}
i.e. $\mathcal{C}$ in \eqref{eq:coherenceTrace} is the same as the squared $\htwo$ norm of the displacement system. (See e.g. \cite{bamieh2012coherence} for the corresponding result for continuous time systems.)

\subsection{The change in coherence from edge additions}
We now turn to the problem of designing edge additions for optimal coherence.
In particular, we show how the coherence changes with the addition of an edge. 
The solution that we present here for discrete time systems differs in several ways from that of continuous time systems reported in the literature \cite{summers2015topology}.
Moreover, our results allow for efficient computation of the improvement in coherence from each possible edge addition in a large scale network.

For the edge addition \eqref{eq:Lapl-A-addition}, we are interested in the difference
\begin{align*}
\mathcal{C}_\delta = \bar{\mathcal{C}} - \mathcal{C},
\end{align*}
where $\bar{\mathcal{C}}$ is the coherence after the edge addition and $\mathcal{C}$ the one before. 

\begin{proposition}
	For any edge addition $\{(\source,\target),w\},\ w > 0$, it holds $\mathcal{C}_\delta \leq 0$.
\end{proposition}
\proof
	As before, let $0 < \lambda_1 \leq \lambda_2 \leq \dots \leq \lambda_{n-1} < \lambda_n = 1$ be the eigenvalues of $A$ and let $0 < \bar{\lambda}_1 \leq \bar{\lambda}_2 \leq \dots \leq \bar{\lambda}_{n-1} < \bar{\lambda}_n = 1$ be the eigenvalues of $\bar{A}$. Since $A - \bar{A} =  e_{\source \target} w e_{\source \target}^\top$ is positive semidefinite, Weyl's eigenvalue inequality applies \cite{horn2012matrix}, giving $\bar{\lambda}_i \leq \lambda_i,\ i=1,\dots,n$. This implies that
	\begin{align*}
	\bar{\mathcal{C}} = \sum_{i=1}^{n-1} \frac{1}{1 - \bar{\lambda}_i^2} \leq \sum_{i=1}^{n-1} \frac{1}{1 - \lambda_i^2} = \mathcal{C},
	\end{align*}
	hence, $\mathcal{C}_\delta = \bar{\mathcal{C}} - \mathcal{C} \leq 0$. 
	 
\qed

According to the previous proposition, the addition of an edge in a consensus network with noise inputs always reduces the coherence $\mathcal{C}$. (Alternatively, in a consensus network with control signal inputs, adding an edge always makes it more difficult to steer the network in the directions of the disagreement subspace since $||\Upsilon_\xi ||_\htwo$ is reduced.)

\begin{theorem}
	The edge addition $\{(\source, \target), w\}$ changes the network coherence by
	\begin{subequations}
		\begin{align}
		\mathcal{C}_\delta 
		&= \frac{\left(e_{\source \target}^\top(I-A^2)^\dagger e_{\source \target}\right)^2}{\alpha_1 \alpha_2} \\
		& \ \ \ \ + \frac{e_{\source \target}^\top(I+A)^{-1} (I-A)^\dagger (I+A)^{-1} e_{\source \target} }{\alpha_1} \nonumber \\
		& \ \ \ \ + \frac{e_{\source \target}^\top(I-A)^\dagger (I+A)^{-1}(I-A)^\dagger e_{\source \target} }{\alpha_2}, 
		\end{align}
		where 
		\begin{align}
		&\alpha_1 = \frac{1}{ w}-e_{\source \target}^\top(I+A)^{-1}e_{\source \target}, \\
		& \alpha_2 = \frac{1}{- w} - e_{\source \target}^\top(I-A)^\dagger e_{\source \target}.
		\end{align}
		\label{eq:deltaCoherenceRes}
	\end{subequations} 	
\end{theorem}

\proof
	From \eqref{eq:coherenceDiscreteEigs}, 
	\begin{align*}
	\bar{\mathcal{C}} &= \trace (I-\bar{A}^2)^\dagger = \trace \left((I-\bar{A})(I + \bar{A})\right)^\dagger \\
	&= \trace  (I + \bar{A})^{-1} (I-\bar{A})^\dagger.
	\end{align*}
	Observe that $I+ \bar{A}$ is non-singular and invertible.
	Apply the matrix inversion lemma \cite{horn2012matrix}, 
	\begin{align*}
	(I+\bar{A})^{-1} &= (I+A - e_{\source \target}  w e_{\source \target}^\top)^{-1} \\
	&= (I+A)^{-1} + \beta_1,
	\end{align*}					 
	where 
	\begin{align*}
	&\beta_1 = (I+A)^{-1} e_{\source \target}  e_{\source \target}^\top(I+A)^{-1}/\alpha_1 
	\end{align*}
	and $\alpha_1$ as in \eqref{eq:deltaCoherenceRes}.
	For the pseudo-inverse $(I-\bar{A})^\dagger$,  
	\cite[Theorem~3]{meyer1973generalized} applies, which gives the similar expression
	\begin{align*}
	(I-\bar{A})^\dagger &= (I-A + e_{\source \target}  w e_{\source \target}^\top)^\dagger \\
	&= (I-A)^\dagger + \beta_2,
	\end{align*}
	where
	\begin{align*}
	& \beta_2 = (I-A)^\dagger e_{\source \target}  e_{\source \target}^\top(I-A)^\dagger/\alpha_2
	\end{align*}
	and $\alpha_2$ as in \eqref{eq:deltaCoherenceRes}. Hence, 
	\begin{align*}
	(I - \bar{A}^2)^\dagger &= \left( (I+A)^{-1} + \beta_1 \right) \left( (I-A)^\dagger + \beta_2 \right) \\
	&= (I+A)^{-1} (I-A)^\dagger + \beta_1 (I-A)^\dagger \\
	&\ \ \ + (I+A)^{-1} \beta_2 + \beta_1 \beta_2 \\
	&= (I-A^2)^\dagger + \beta_1 (I-A)^\dagger + (I+A)^{-1} \beta_2 + \beta_1 \beta_2, 
	\end{align*}
	which implies that
	\begin{align*}
	\mathcal{C}_\delta &= \bar{\mathcal{C}} - \mathcal{C} = \trace(I-\bar{A}^2)^\dagger - \trace(I-A^2)^\dagger \\
	&=\trace \left( \beta_1 (I-A)^\dagger + (I+A)^{-1}\beta_2 + \beta_1 \beta_2 \right). 
	\end{align*}
	It is straight-forward to verify that this expression evaluates to \eqref{eq:deltaCoherenceRes} using the cyclic property of the trace operator.  
\qed

%

The next proposition is straightforward and is used to obtain an efficient computation of \eqref{eq:deltaCoherenceRes} for all possible edge additions $\{(\source, \target), w\},\  \source,\target = 1,\dots,n,\ \source \neq \target$, and $w\in \mathbb{R}^+$ given.
\begin{proposition}
	Let $M\in \mathbb{R}^{n\times n}$ and symmetric. Then 
	\begin{align}
	N = \onevec \diag(M)^\top + \diag(M) \onevec^\top - 2 M %
	\end{align}
	is such that
	\begin{align*}
	e_{ij}^\top M e_{ij} = N_{ij}.
	\end{align*}
	\label{prop:matrElementFromEij}
\end{proposition}
Given the factorization $A = U\Lambda U$, all the matrices 
\begin{align*}
&(I-A)^\dagger, (I+A)^{-1},	(I-A^2)^\dagger, \\
&(I+A)^{-1}(I-A)^\dagger (I+A)^{-1}, \text{ and } \\
&(I-A)^\dagger (I+A)^{-1}(I-A)^\dagger
\end{align*}
appearing in the right hand side of \eqref{eq:deltaCoherenceRes} can be easily computed with elementary operations (see e.g. \eqref{eq:ImAsqPinv} for $(I-A^2)^\dagger$).
By applying Proposition \ref{prop:matrElementFromEij} to these matrices, a matrix $Q$ can be constructed with elements $Q_{\target \source} = \mathcal{C}_\delta$ for the edge addition $\{(\source, \target), w\}$.


\section{Applications}
\label{sec:applications}

\subsection{Edge modifications and the degree of controllability}

For internally stable networks, edge modifications can be used as a mean to improve the degree of controllability \cite{becker2017network,chanekar2019}. 
One metric for the energy needed for control is $\trace(W)$, or $\trace(CWC^\top)$ when only the states of certain output nodes are considered.
In this context, equation \eqref{eq:positiveLowerBound} provides a lower bound on the increment that the edge addition $\{(s,\target),w\},\ 0<w<\gainm$, gives to the trace of the Gramian of a positive network. 
With $W$ the controllability Gramian for $(A,B)$ and $\bar{W}$ the controllability Gramian for $(\bar{A},B)$, it is
\begin{align*}
	\trace \left(C \bar{W}C^\top \right) &= ||\tf{\bar{A}}{\outputs \drivers}||_\htwo^2 = ||\tf{A}{\outputs \drivers}  + G^\delta||_\htwo^2 \\
	&\geq ||\tf{A}{\outputs \drivers}||_\htwo^2 + ||G^\delta||_\htwo^2\\
	&\geq \trace \left(C W C^\top \right) + \frac{p_\target w^2 q_\source}{1- \varepsilon_{\target\to \source}w^2}.
\end{align*}

One way to improve $\trace \left(C WC^\top \right)$ is therefore to make edge modifications in a greedy manner, choosing the edges that correspond to the largest bounds \eqref{eq:positiveLowerBound}.

\subsection{Network fragility}

Fragility of internally stable networks can be defined in many different ways.
In \cite{pasqualetti2018fragility}, fragility refers to the sensitivity of a network to variations in the edge weights and it is quantified by the stability radius, 
\begin{align*}
r(A) = \min \{\bar{\sigma}(\Delta) \text{ s.t. } \rho(A+ \Delta) \geq 1 \},
\end{align*} 
i.e. the spectral norm of the smallest change in the network weights that renders it unstable. 
\cite{pasqualetti2018fragility} assumes no particular structure on the matrix $\Delta$. However, if we restrict $\Delta$ to the set of real matrices with only one non-zero entry, then it represents an edge modification as studied in this paper. 
For this case the metric $r(A)$ is in fact the stability margin $\gainm$ of Theorem \ref{prop:stabRad}.

\subsection{Simulations}

\begin{figure}
	\centering
	\includegraphics[width=0.9\columnwidth]{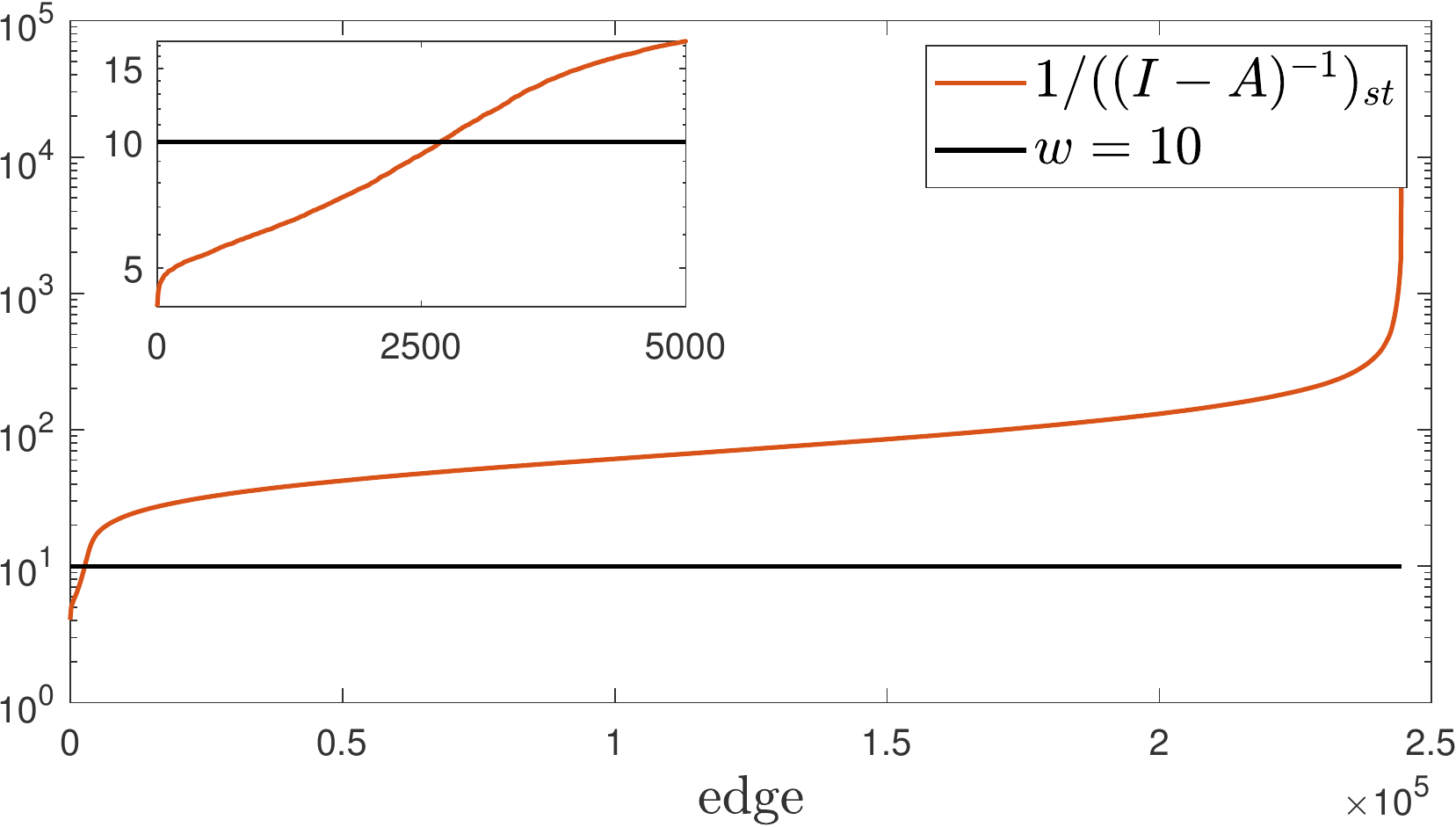}
	
	(a)
	
	\includegraphics[width=0.9\columnwidth]{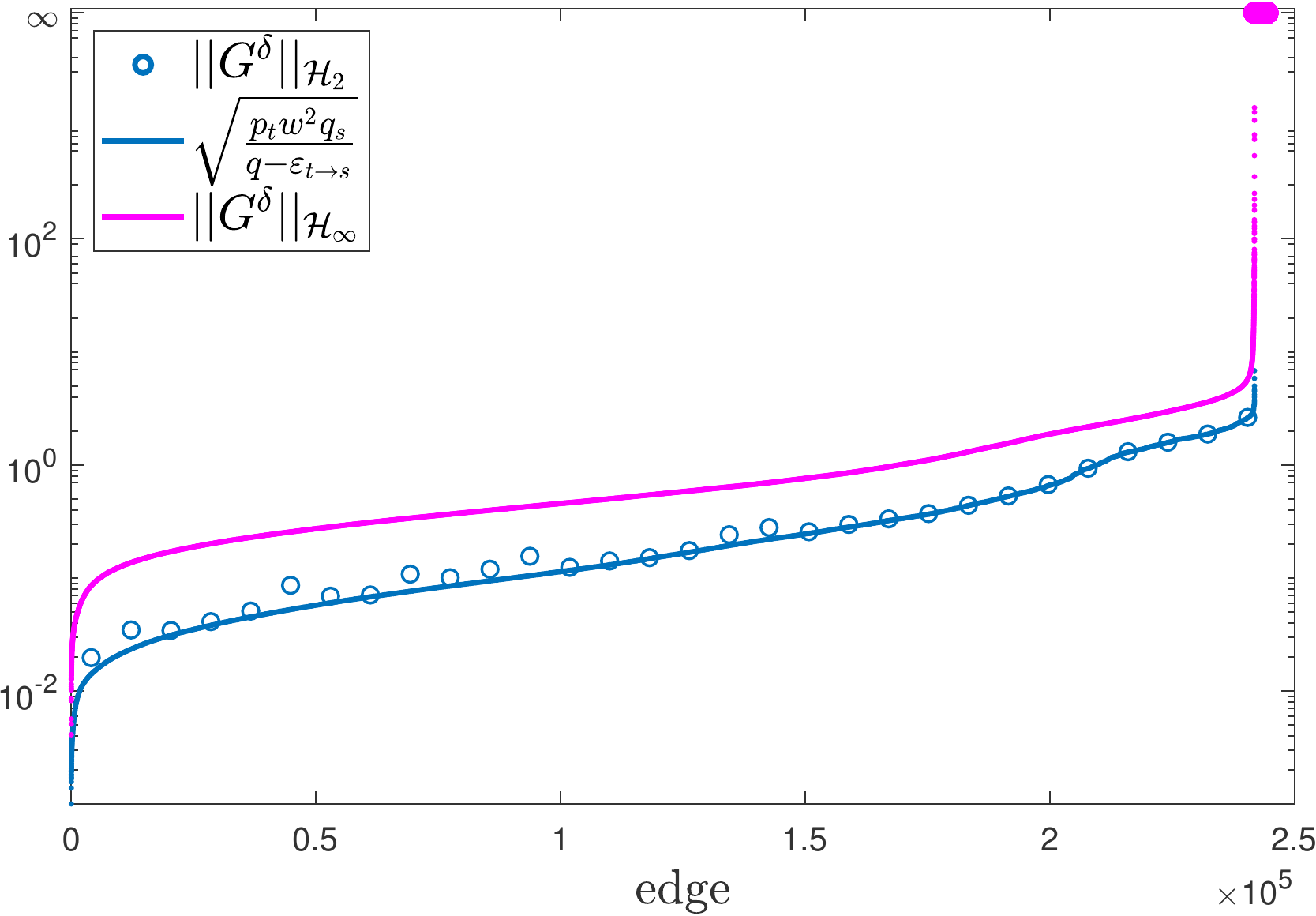}
	
	(b)
	
	\caption{Numerical computations on a random \ErdosRenyi network with 500 nodes. For all $\source,\target \in \mathcal{V}$, $\source \neq \target$ and weight $w=10$, the plots show
	(a) the stability margins $\gainm$, 
	(b) the norm $||G^\delta||_\hinf$ and the lower bound of $||G^\delta||_\htwo$. Edges are ordered along the x-axis in ascending order in both cases. For a sub-selection of 30 edges, also the exact value of $||G^\delta||_\htwo$ has been computed. }
	\label{fig:gm}
\end{figure}

The metrics for the impact of edge modifications of Section \ref{sec:stableDynamics}
are computed for a random \ErdosRenyi network with 500 nodes and plotted in Figure \ref{fig:gm}. Edges are generated with probability $0.02$ and with weights that are first sampled from the uniform distribution over $]0\ 1]$, and then rescaled such that the prespecified spectral radius $\rho(A) = 0.9$ is obtained. (Hence, the network is positive and stable.) $50$ input nodes and $100$ output nodes are randomly selected.
In Figure \ref{fig:gm}(a), the stability margins $\gainm$ are plotted for all $\source,\target \in \mathcal{V},\ \source \neq \target$. 
The figure also shows the weight $w = 10$. 
For the $\approx 2500$ edges $\{(\source,\target),w\}$ at the left end of the plot (see the inset), it holds $w \geq \gainm$. Hence, in these cases the modification $\{(\source,\target),w\}$ renders the network unstable.
The metric $||G^\delta||_\hinf$ is computed and plotted in Figure \ref{fig:gm}(b) for each single edge modification $\{(\source,\target),w\}$, $\source,\target \in \mathcal{V},\ \source \neq \target$ and $w = 10$. 
The modifications that result in instability can be observed here as marks at the top right corner of the figure.
The lower bound \eqref{eq:positiveLowerBound} on $||G^\delta||_\htwo$ is also plotted for each possible single edge modification. For comparison, the exact values are computed for a few edges using standard Matlab routines. (This is however infeasible to do for all edges due to the computational cost.) It appears that the bound is consistently close to the exact value in the cases where it has been computed.

\begin{figure}
	\centering
	\includegraphics[width=.7\columnwidth]{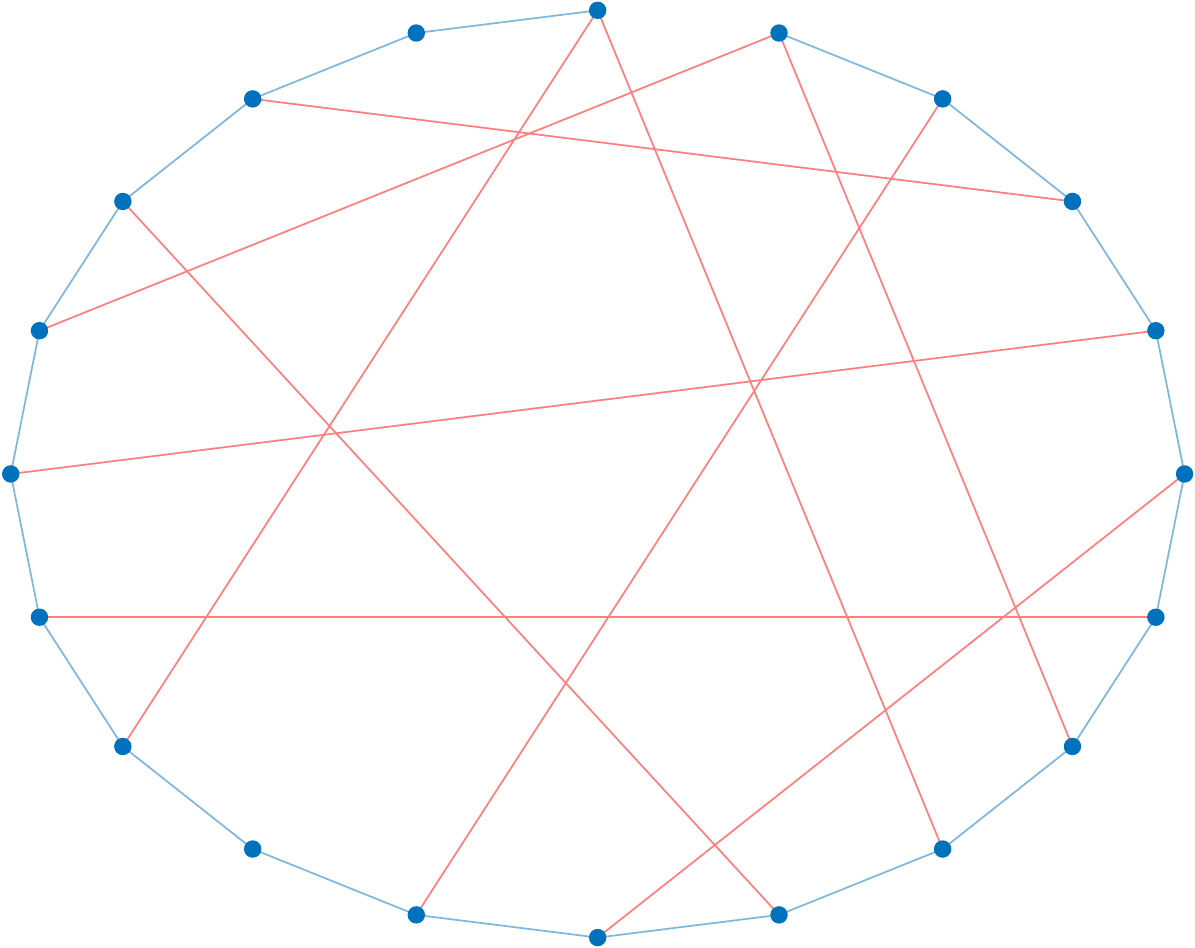}
	\caption{Iterative edge addition in networks with Laplacian dynamics. The original (blue) line network is grown with new edges (red) that efficiently improve the network coherence. }
	\label{fig:iterativeEdgeAddition}
\end{figure}

Figure \ref{fig:iterativeEdgeAddition} presents an example of iterative edge addition to optimize the coherence in networks with Laplacian dynamics and subject to noise inputs on each node.
Starting with a line network (blue), edges are added one at a time (red), each time selecting the edge that gives the largest improvement $\mathcal{C}_\delta$. All edges have weight $0.2$. When 10 edges have been added, the coherence has reduced from $\mathcal{C} = 173.37$ to $30.8$. It appears that edges are added in a way that efficiently reduces the shortest path between any pair of nodes (from 19 in the original network to 4 in the grown network).
While this example is of small scale to illustrate how edges are added, our method is feasible also for large scale networks with arbitrary topology.

\section{Conclusions}

In large scale networks with stable dynamics, the particular structure of the transfer function $G^\delta$ that we derive for the changes in network output due to an edge modification enables us to quantify the impact of each possible edge modification.
Whether we use the $\htwo$ norm or the $\hinf$ norm as metric, the impact from modifying the edge $(\source,\target)$ depends on three network properties: (i) the strength of the connections from the input nodes to $\source$, (ii) that from $\target$ to the output nodes, and (iii) the feedback connections from $\target$ to $\source$.
In particular, the third factor appears not to have been observed before in the context of network control systems. 
In the case of stable dynamics it provides a stability margin which leads to an upper bound on the admissible edge weights.

When Laplacian dynamics is imposed, the third factor becomes less important, as the Laplacian structure takes care of automatically maintaining the state update matrix at the marginal stability boundary.

As a possible application of our results, we show how the proposed $\htwo$ metric can be used to design edge modifications that improve the trace of the controllability Gramian control energy metric.
Moreover, the stability margins we present for networks subject to edge modifications has possible applications to network robustness/fragility.
In this case, an interesting extension of our results would be for instance to consider the stability margins when $k \in 2,3,\dots$ arbitrary edges are allowed to be modified simultaneously, rather than one at a time.

\bibliographystyle{abbrv}


\end{document}